\documentstyle[epsfig,preprint,aps,eqsecnum,fixes]{revtex}

\makeatletter           
\@floatstrue
\def\figure{\let\@capwidth\columnwidth\@float{figure}}
\let\endfigure\end@float
\@namedef{figure*}{\let\@capwidth\textwidth\@dblfloat{figure}}
\@namedef{endfigure*}{\end@dblfloat}
\def\table{\let\@capwidth\columnwidth\@float{table}}
\let\endtable\end@float
\@namedef{table*}{\let\@capwidth\textwidth\@dblfloat{table}}
\@namedef{endtable*}{\end@dblfloat}

\makeatother

\tightenlines
\begin {document}

\def\Tomasik{Tom\'{a}\v{s}ik}
\def\Np{N_{\rm p}}
\def\etal{{\it et al.}}
\def\nc{n_{\rm c}}
\def\c{{\rm c}}
\def\phiphic{\langle\phi^2\rangle_\c}
\def\infmu{{\delta\mu}}

\def\dtmo{{{d\over2}-1}}
\def\mucone{\muc^{[1]}}
\def\bball{{\rm bball}}
\def\dn{\delta n}
\def\dnbball{\dn_\bball}
\def\sunset{{\rm sun}}
\def\tomega{\omega^+}
\def\tPi{\Pi^+}
\def\blparen{\mbox{\boldmath$($}}
\def\brparen{\mbox{\boldmath$)$}}
\def\Mir{{\cal M}}
\def\Nc{N_{\rm c}}

\def\Tc{T_{\rm c}}
\def\kB{k_{\rm B}}

\def\grad{\mbox{\boldmath$\nabla$}}
\def\p{{\bf p}}
\def\q{{\bf q}}
\def\k{{\bf k}}
\def\l{{\bf l}}

\def\half{{\textstyle{1\over2}}}

\def\Li{{\rm Li}}

\def\c{{\rm c}}

\def\threehalf{{\textstyle{3\over2}}}

\def\x{{\bf x}}
\def\muc{\mu_{\rm c}}

\def\SI{S_{\rm I}}
\def\eps{\epsilon}
\def\msb{{\overline{\rm MS}}}
\def\msbar{{\overline{\scriptscriptstyle{\rm MS}}}}
\def\rms{r_\msbar^{}}
\def\MSbar{$\msb$}
\def\Mren{M}
\def\bare{{\rm bare}}
\def\gammaE{{\gamma_{\scriptscriptstyle{\rm E}}}}
\def\sumint{\hbox{$\sum$}\!\!\!\!\!\!\int}
\def\PP{{\rm P.P.}}


\preprint {UW/PT--01--19}

\title
{$\Tc$ for homogeneous dilute Bose gases: a second-order result}

\author {Peter Arnold$^a$, Guy Moore$^b$, and Boris \Tomasik$^a$}
\address
    {%
    ${}^a$
    Department of Physics,
    University of Virginia,
    P.O. Box 400714 \\
    Charlottesville, Virginia  22904-4714
    }%
\address
    {%
    ${}^b$
    Department of Physics,
    University of Washington,
    Seattle, Washington 98195
    }%
\date{\today}

\maketitle
\vskip -20pt

\begin {abstract}%
{%
   The transition temperature for a dilute, homogeneous, three-dimensional
   Bose gas has the expansion
   $\Tc = T_0 \{1 + c_1 a n^{1/3} + [c_2' \ln(a n^{1/3}) + c_2''] a^2 n^{2/3}
    + O(a^3 n)\}$,
   where $a$ is the scattering length, $n$ the number density, and $T_0$
   the ideal gas result.
   The first-order coefficient $c_1$ depends on non-perturbative physics.
   In this paper, we show that the coefficient $c_2'$ can be computed
   perturbatively.  We also show that the remaining second-order coefficient
   $c_2''$ depends on non-perturbative physics but can be related, by a
   perturbative calculation, to quantities that have previously
   been measured using lattice simulations of three-dimensional O(2)
   scalar field theory.
   Making use of those simulation results, we find
   $\Tc \simeq
    T_0 \{1 + (1.32\pm0.02) \, a n^{1/3}
    + [19.7518 \ln(a n^{1/3}) + (75.7\pm0.4)] a^2 n^{2/3}
    + O(a^3 n)\}$.
}%
\end {abstract}

\thispagestyle{empty}


\section {Introduction}

Long-distance physics at a second-order phase transition is generically
non-perturbative.  For this reason, researchers have found it non-trivial
to compute corrections to the ideal gas result for the critical temperature
$\Tc$ for Bose-Einstein condensation (BEC) of a dilute, homogeneous Bose gas
in three dimensions.  It is currently understood that the correction
$\Delta\Tc \equiv \Tc - T_0$ to the ideal gas result $T_0$
behaves parametrically as
\begin {equation}
   {\Delta\Tc\over T_0} \to c_1 a n^{1/3}
\label {eq:LOform}
\end {equation}
in the dilute (or, equivalently, weak-interaction) limit, where
$c_1$ is a numerical constant,
$a$ is the
scattering length, which parameterizes the low energy 2-particle
scattering cross-section, and $n$ is the density of the homogeneous
gas.
We will assume that the interactions are repulsive ($a > 0$).
A clean argument for (\ref{eq:LOform}) may be found in
Ref.\ \cite{baym1}, which also shows how the problem of calculating
the constant $c_1$ can be reduced to a problem in three-dimensional
O(2) field theory.  Recent numerical simulations of that theory have
obtained the results $c_1 = 1.29 \pm 0.05$ \cite{prokofev} and
$c_1 = 1.32 \pm 0.02$ \cite{boselat1,boselat2}.

In this paper, we shall extend the result for $\Tc(n)$ to second order in $a$
for a homogeneous Bose gas.
This is also the relationship between $\Tc$ and the central number density
for a Bose gas in an arbitrarily wide trap.
(In contrast, the relationship between $\Tc$ and the total number $\Np$ of
particles in a trap depends on somewhat different
physics.  A second-order result for $\Tc(\Np)$ in an arbitrarily wide trap
may be found in Ref.\ \cite{trap}.)

In the homogeneous case,
Holzmann, Baym, and Lalo\"e \cite{logs}
have recently argued that a logarithmic term
appears at second order,
\begin {equation}
   {\Delta\Tc\over T_0} \longrightarrow
          c_1 a n^{1/3} + c_2' a^2 n^{2/3} \ln(a n^{1/3}) + \cdots ,
\end {equation}
and they made a rough estimate of the coefficient $c_2'$ using large-$N$
arguments.  A similar logarithm has been found for $\Tc(\Np)$
in the case of trapped gases \cite{trap}.
We will show that, in contrast to $c_1$, the coefficient
$c_2'$ of the logarithm can be computed exactly using perturbation theory.
Our result is
\begin {equation}
   c_2' = - {64\pi\,\zeta(\half) \over 3 \left[\zeta(\threehalf)\right]^{5/3}} 
   \, ,
\end {equation}
where $\zeta(s)$ is the Riemann zeta function.
We will compute this result by performing a second-order
perturbative calculation to
match the physics of the transition onto that of three-dimensional
O(2) scalar field theory.  The same matching calculation will also
perturbatively determine the relationship between the {\it non}-logarithmic
term at second order and certain non-perturbative quantities in O(2) scalar
theory which have been previously measured in lattice simulations.
As a result, we will determine all the coefficients in the
second-order expansion
\begin {equation}
   {\Delta\Tc\over T_0} =
          c_1 a n^{1/3}
          + \left[c_2' \ln(a n^{1/3}) + c_2''\right] a^2 n^{2/3} +
          O(a^3 n) \, ,
\label {eq:Tcform}
\end {equation}
[where the notation $O(a^3 n)$ is not intended to make any particular claim
about what powers of logarithms might appear at third order].

We should emphasize that when we refer to the ``first order'' and
``second order'' terms in (\ref{eq:Tcform}), we do not mean first
and second order in perturbation theory.  Perturbation theory breaks down
for these quantities, and that breakdown manifests as the appearance of
infrared (IR) infinities beyond a certain order in the perturbative expansion.

A portion of the required perturbative matching calculations has already been
performed in Ref.\ \cite{trap}, which also gives a discussion of the
philosophy and methods of perturbative matching calculations between Bose
gases and three-dimensional O(2) field theory at the phase transition.
However, we find it convenient to use slightly different conventions than
Ref.\ \cite{trap}.  For the sake of introducing conventions and
notation, and for the sake of making this article somewhat self-contained, we
will use the remainder of this introduction to briefly review matching and the
different distance scales associated with physics at the transition.
In section \ref{sec:UV}, we will fix the ultraviolet (UV) regularization and
renormalization schemes we will use for our calculation.
We then proceed to do the matching calculations and
assemble all the matching results in section \ref{sec:match},
though we leave the details of the more
intricate diagrammatic calculations for later sections.
In section \ref{sec:final}, we then put together our final results for
the relationship between $T$ and $n$ at the transition.
Section \ref{sec:bball} gives the details of how to calculate the most
complicated diagram that was needed for matching.
Section \ref{sec:sunset} reproduces some previous results related
to the critical value $\muc(T)$ of the chemical potential \cite{trap}
in a new form that is needed for our analysis.
Finally, section \ref{sec:largeN} explains how our result for the
second-order logarithm is modified in theories with $N$ fields, for the
sake of readers who may wish to compare exact results to the
approximate large-$N$ analysis of Holzmann, Baym, and Lalo\"e \cite{logs}.
A very brief outline of how to calculate a few simple finite-temperature
integrals in dimensional regularization is left for an appendix.

For the calculations in this paper, it will be convenient to follow
Baym \etal\ \cite{baym1} and calculate the critical density $\nc(T)$ as
a function of $T$ rather than the critical temperature $\Tc(n)$ as a
function of $n$.  We will then obtain the formula for $\Tc(n)$ by
inverting the relationship.  The ideal gas result $n_0(T)$ for $\nc$ is
\begin {equation}
    n_0(T) = {\zeta(\threehalf)\over\lambda^3} ,
\end {equation}
where
\begin {equation}
   \lambda \equiv \sqrt{2\pi\over mT}
\label {eq:lambda}
\end {equation}
is the thermal wavelength.
(In this paper, we work in units where $\hbar=\kB=1$, where $\kB$ is
Boltzmann's constant.)
The diluteness condition
$a n^{1/3} \ll 1$ for the expansions discussed above can therefore
alternatively be expressed as
\begin {equation}
   a \ll \lambda
\label {eq:heirarchy}
\end {equation}
at the transition.


\subsection {Overview of matching to 3-dimensional O(2) theory}

Baym \etal\ \cite{baym1}
were the first to use an effective three-dimensional O(2)
scalar field theory to study non-universal long-distance physics
of the BEC transition of a dilute Bose gas.
A more systematic discussion of how to match the parameters of the
O(2) theory to the original problem, in order to study interaction
effects beyond leading order, may be found in Ref.\ \cite{trap}.
Here, we will briefly review these issues in preparation for doing
the matching calculations that we will need to obtain $\Tc(n)$ at second
order.  Perturbative matching calculations, which allow effective theories
to be used to calculate non-universal quantities, can be performed whenever
the {\it short}-distance physics described by the effective theory is
perturbative, at a scale where the effective theory is still applicable.
Such calculations have a long history that includes
lattice field theory \cite{symanzik},
Bose condensation
at zero temperature \cite{Bose},
relativistic corrections to non-relativistic QED \cite{QED},
heavy quark physics \cite{heavy quarks},
ultra-relativistic plasmas \cite {Braaten&Nieto},
and non-relativistic plasma physics \cite{nonrel plasma}.
For a general discussion, see also Ref. \cite{effective}.

The starting point is the well-known description of a dilute Bose gas
by a second-quantized Schr\"odinger equation,
together with a chemical potential
$\mu$ that couples to particle number density $\psi^*\psi$, and a
$|\psi|^4$ contact interaction that reproduces low-energy scattering
\cite {review}.
The corresponding Lagrangian is
\begin {equation}
   {\cal L} = \psi^* \left(
        i \, \partial_t + {1\over 2m} \, \nabla^2
        + \mu \right) \psi
    - {2\pi a\over m} \, (\psi^* \psi)^2 .
\label {eq:L1}
\end {equation}
This effective description is valid for distance scales large compared to
the scattering length $a$.
Corrections to this description,
due to the energy dependence of the cross-section
or 3-body interactions or so forth,
do not affect $\Tc(n)$ at second order
(see section \ref{sec:final}).
At finite temperature, it is convenient to study (\ref{eq:L1})
using the imaginary time formalism, in which
$t$ becomes $-i\tau$ and imaginary time $\tau$
is periodic with period $\beta = 1/T$.
The imaginary-time action is then
\begin {equation}
   \SI = \int_0^{\beta} d\tau \int d^3x \left[
        \psi^* \left(
        \partial_\tau - {1\over 2m} \, \nabla^2
        - \mu \right) \psi
    + {2\pi a\over m} \, (\psi^* \psi)^2
    \right] .
\label {eq:SI}
\end {equation}
We shall call this the 3+1 dimensional theory, referring to three spatial
dimensions plus one (imaginary) time dimension.  The expectation value of
the number density is
given by
\begin {equation}
    n = \langle \psi^* \psi \rangle .
\end {equation}

As usual,
the field $\psi$ can be decomposed into imaginary-time
frequency modes $\psi_j$ with discrete Matsubara
frequencies $\omega_j = 2\pi j/\beta$, where $j$ is an integer.
If we ignore interactions for a moment, and treat $\mu$ as small,
then non-zero Matsubara frequency modes in (\ref{eq:SI}) are associated
with a correlation length of order $(2m\omega_j)^{-1/2} \lesssim \lambda$,
where $\lambda$ is the thermal wavelength (\ref{eq:lambda}).
Near the transition, at distance scales large compared to the thermal
wavelength $\lambda$,
all the modes with non-zero Matsubara frequencies decouple, leaving behind
an effective theory of just the zero-frequency modes $\psi_0(\x)$.
If one were to naively throw away the non-zero frequency modes from
the original 3+1 dimensional action (\ref{eq:SI}), it would reduce to
\begin {equation}
   \SI \to \beta \int d^3x \left[
        \psi_0^* \left(
        - {1\over 2m} \, \nabla^2
        - \mu \right) \psi_0
    + {2\pi a\over m} \, (\psi_0^* \psi_0)^2
    \right] .
\label {eq:S3naive}
\end {equation}
This is the rough form of the effective 3-dimensional theory of $\psi_0$,
which describes long-distance physics at the transition for time-independent
quantities.  However, completely ignoring the effects of non-zero frequency
modes was an oversimplification.  In field theories, the
short-distance and/or
high-frequency modes do have effects on long-distance physics, but those
effects can be absorbed into (1) a modification of the strengths of
relevant interactions (in the sense of the renormalization group)
between the long-distance/zero-frequency fields, and
(2) the appearance of additional
marginal and irrelevant interactions between the
long-distance/zero-frequency modes.
The latter effect will not be relevant at second order for $\Tc(n)$
(see section \ref{sec:final}).  Because of the first effect,
the correct three-dimensional effective theory is of the more general form
\begin {equation}
   S_3 = \beta \int d^3x \left[
        \psi_0^* \left(
        - {Z_\psi \over 2m} \, \nabla^2
        - \mu_3 \right) \psi_0
    + Z_a {2\pi a\over m} \, (\psi_0^* \psi_0)^2
    + f_3
    \right] ,
\label {eq:S3}
\end {equation}
where the difference of the ($\psi_0$-independent) parameters
$Z_\psi$, $\mu_3$, $Z_a$, and $f_3$ from
the naive values $Z_\psi=Z_a=1$, $\mu_3=\mu$, and $f_3=0$
of (\ref{eq:S3naive}) incorporates the
effects of short-distance physics on long-distance physics.
These parameters can be computed perturbatively because short-distance physics
is perturbative.
The above three-dimensional theory is super-renormalizable and has
UV divergences associated with the parameters $\mu_3$ and $f_3$.
The coefficients of the other terms, however, have a simple finite relationship
to the parameters of the original theory.
The parameter $f_3$
represents the contributions of the non-zero Matsubara frequency modes
to the free energy density (along with any associated UV counterterms of the
three-dimensional theory).

It is often conventional to rescale the field
of the effective
three-dimensional theory (\ref{eq:S3}) as%
\footnote{
  The rescaling used here differs from that of Ref.\ \cite{trap} by a factor
  of $Z_\psi^{-1/2}$.  This difference of convention is actually moot since
  $Z_\psi=1$ at second order.
}
\begin {equation}
  \psi = \left(m T \over Z_\psi\right)^{1/2} (\phi_1 + i \phi_2)
\end {equation}
and write
\begin {equation}
   S_3 = \int d^3x \left[
        {1\over 2} \, |\grad\phi|^2
        + {r_\bare\over 2} \, \phi^2
        + {u \over 4!} \, (\phi^2)^2
    + {\cal F}_3
    \right] ,
\label {eq:O2}
\end {equation}
where $\phi=(\phi_1,\phi_2)$ is a real 2-vector,
$\phi^2 \equiv \phi_1^2+\phi_2^2$, and
\begin {equation}
   r_\bare = - {2m \mu_3 \over Z_\psi},
   \qquad
   u = {96\pi^2 a\over\lambda^2} \, {Z_a\over Z_\psi^2}\, ,
   \qquad
   {\cal F}_3 = \beta f_3 .
\label {eq:ru}
\end {equation}
This is O(2) scalar field theory in three dimensions.
This form makes it easy to understand the scale at which physics becomes
non-perturbative.
The only parameters of the $\phi$-dependent part of $S_3$ are
$r_\bare$ and $u$.
For fixed $u$, 
imagine finding the $r_\bare(u)$ that corresponds to
the phase transition.  Then all correlations at the phase transition
can be considered as determined by $u$.
By dimensional analysis, the distance scale of non-perturbative physics
is therefore $1/u$.
$Z_a$ and $Z_\psi$ turn out to be perturbatively close to 1, and so
this scale is $1/u \sim \lambda^2/a$ by (\ref{eq:ru}).

Perturbation theory in the three-dimensional theory (\ref{eq:O2})
is an expansion in $u$.
Consider the dimensionless cost of each order of perturbation theory.
By dimensional analysis, the contribution to that cost by physics at
a momentum scale of order $p$ must be order $u/p$.
This means that perturbation theory
breaks down for distance scales $p^{-1} \gtrsim u^{-1}$.
It also means that, thanks to (\ref{eq:heirarchy}), perturbation theory
works fine for distance scales $p^{-1} \lesssim \lambda$ at the transition,
which are the distance
scales of the non-zero Matsubara frequency modes.
This is the reason that those modes can be
treated perturbatively and a perturbative matching calculation is possible.

To compute the number
density at a given temperature and chemical potential, it is convenient to
rewrite $n = \langle\psi^*\psi\rangle$ as
\begin {equation}
   n = (\beta V)^{-1} {\partial\over\partial\mu} \, \ln Z
     = - (\beta V)^{-1}
          \left\langle \partial \SI \over \partial \mu\right\rangle .
\end {equation}
In the equivalent three-dimensional description (\ref{eq:S3}), this becomes
\begin {eqnarray}
   n &\simeq& - (\beta V)^{-1}
          \left\langle\partial S_3 \over \partial\mu \right\rangle
\nonumber\\
     &=& 
       - {\partial Z_\psi \over\partial\mu} \, {1\over2m}\,
               \langle|\grad\psi_0|^2 \rangle
       + {\partial\mu_3\over\partial\mu} \, \langle \psi_0^* \psi_0 \rangle
       - {\partial Z_a\over\partial\mu} \, {2\pi a\over m} \,
                \langle (\psi_0^* \psi_0)^2 \rangle
       - {\partial f_3\over \partial\mu} \,,
\label {eq:nmatch0}
\end {eqnarray}
where all the expectations are taken in the purely three-dimensional
theory (and UV-regularized, as necessary).
The only approximation made here is ignoring corrections that would
appear as higher-dimensional interactions in (\ref{eq:S3}) which,
as we've already said
we will review later, do not affect the calculation of $\Tc$ at second
order.
Also, we shall see that
the derivative $\langle |\grad\phi|^2 \rangle$ term and
the quartic $\langle (\psi_0^* \psi_0)^2 \rangle$ term
are not relevant at second order, so that one may simply take
\begin {equation}
   n \simeq {\partial\mu_3\over\partial\mu} \, \langle \psi_0^* \psi_0 \rangle
       - {\partial f_3\over \partial\mu} \,.
\label {eq:nmatch}
\end {equation}

A very similarly structured matching calculation was undertaken in
Ref.\ \cite{boselat2} for matching the continuum 3-dimensional theory to
a lattice 3-dimensional theory, where the purpose was to match the
two theories order by order in the lattice spacing, in order to improve
approach to the continuum limit in numerical simulations.
The structure of that calculation,
and the topology of the required perturbative diagrams,
is identical to what we will need for the present task.
The only substantial difference is that we will be evaluating those diagrams in
the 3+1 dimensional theory, rather than 3-dimensional lattice theory,
and that the perturbative expansion will correspond to an expansion in the
scattering length (or $a n^{1/3}$) rather than the lattice spacing.
There is additionally a trivial difference in presentation:
Ref.\ \cite{boselat2} did
not keep track of $\phi$-independent terms analogous to $f_3$, which we
use in (\ref{eq:nmatch}), but instead
discussed the matching of $\psi^* \psi$ directly.

The generic technology of matching calculations was reviewed in the context of
our current problem in section IV.A of Ref.\ \cite{trap}.  The idea is to
perturbatively calculate an identical finite set of physical infrared
quantities in the 3+1 dimensional and 3 dimensional theories,
and then equate the
answers to determine the parameters of the 3 dimensional effective theory.
Each of the calculations must be IR-regulated, but the dependence on the
choice of IR regulator will disappear in the final result of the matching
(provided the same IR regulator is used for both theories).
In our case, this is because matching is accounting for the differences of
the two theories at the short distance scales ($\lesssim\lambda$) associated
with the non-zero Matsubara frequency modes.
For the specific purpose of the matching calculation,
$\mu$ can be formally treated as a perturbation, along with the quartic
interaction proportional to $a$.
That is because
the short distance scales $\lesssim\lambda$ are associated
with energies per particle $\gtrsim T$, which is large compared to the
chemical potential $\mu$ at (or very near) the transition.
With $\mu$ treated perturbatively, the imaginary time Feynman rules are
given in Table \ref{tab:feyn}.
We will use the notation $k_0$, $l_0$, $p_0$, ... to designate the
Matsubara (imaginary time) frequencies associated with propagators with
momenta $\k$, $\l$, $\p$, ..., and have introduced the short-hand notation
\begin {equation}
       \omega_k \equiv {k^2\over 2m} \,.
\end {equation}

\begin{table}[t]
\begin {center}
\tabcolsep=8pt

\begin {tabular}{|c|c|c|}             \hline
  & 3+1 dim.\ theory of $\psi$
  & 3 dim.\ theory of $\psi_0$
\\ \hline
  \centering{$\vcenter{\hbox{\epsfig{file=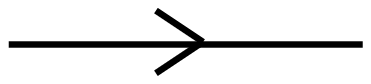,scale=.3}}}$}
  & $\displaystyle{1\over i k_0 + \omega_\k}$
  & $\displaystyle{Z_\psi^{-1}  \over \omega_\k}$
\\
  \centering{$\vcenter{\hbox{\epsfig{file=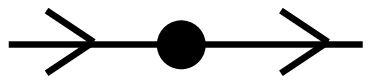,scale=.3}}}$}
  & $+\mu$
  & $+\mu_3$
\\
  \centering{$\vcenter{\hbox{\epsfig{file=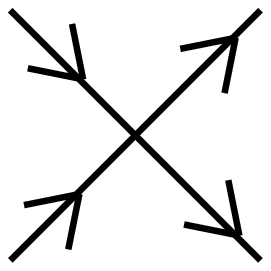,scale=.3}}}$}
  & $\displaystyle{-{8\pi a\over m}}$
  & $\displaystyle{-Z_a\,{8\pi a\over m}}$
\\ \hline
\end {tabular}
\end {center}
\caption
    {%
    \label {tab:feyn}
       Feynman rules, appropriate for a matching calculation in a uniform Bose
       gas, for the original 3+1 dimensional theory (\ref{eq:SI}) of $\psi$
       and the effective three-dimensional theory (\ref{eq:S3}) of $\psi_0$.
       The variable $k_0$ represents the Matsubara frequency of the field,
       while $\omega_k \equiv k^2/2m$.
       At finite temperature, loop frequencies $l_0$ are summed over the
       discrete values $l_0 = 2\pi n T$ with $n$ any integer.
       In dimensional regularization with the \MSbar\ renormalization scheme,
       a factor of
       $\Mren^\eps = (e^{\gammaE/2}\bar\Mren/\sqrt{4\pi})^\eps$ should also be
       associated with each 4-point vertex but has not been explicitly shown
       above.
    }%
\end{table}

\begin {figure}
\vbox{
   \begin {center}
      \epsfig{file=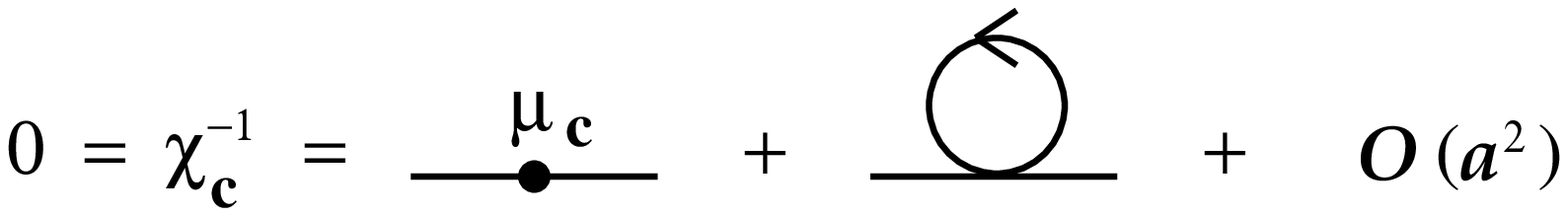,scale=.7}
   \end {center}
   \caption{
       The vanishing of the inverse susceptibility at the phase transition,
       expressed in terms of diagrams at first order in $a$.
       \label {fig:chi1}
   }
}
\end {figure}

In order to streamline calculations later on, it is useful to review the
fact that the critical value $\muc(T)$ of chemical potential has a
somewhat different dependence on non-perturbative physics than the
critical value $\nc(T)$ of the density.  Whereas $\nc(T)$ becomes
non-perturbative at first order in interactions, non-perturbative
effects do not enter the calculation of $\muc(T)$ until
second order.  (See Ref.\ \cite{trap} for a discussion.)
To determine $\muc(T)$ to first order, it is adequate to do a purely
perturbative calculation directly in the original 3+1 dimensional theory.
In particular, one can calculate the inverse susceptibility $\chi^{-1}$
of $\psi$ and set it to zero to determine the transition point, as in
Fig.\ \ref{fig:chi1}.  Such a purely perturbative calculation would be
inadequate at second order, for which one can instead marry perturbative
matching calculations with non-perturbative results from the 3 dimensional
theory \cite{trap}.

\section {UV regularization}
\label {sec:UV}

Before starting a detailed calculation, we have to choose our convention for
regulating ultraviolet divergences in our 3+1 and 3 dimensional effective
theories.  We will use dimensional regularization, replacing the 3 spatial
dimensions by $d=3-\eps$ dimensions.  One convenience of this choice is that,
in the 3+1 dimensional theory, loop corrections to the zero-energy $2\to2$
scattering amplitude vanish at zero temperature and zero chemical potential,
so that there are no corrections to the identification of the $a$ in
(\ref{eq:SI}) with the scattering length \cite{eric}.

We will define UV-renormalized
parameters using the modified minimum subtraction
(\MSbar) scheme,
and we shall call the associated renormalization momentum scale
$\bar\Mren$.  At second order, this will not require any UV subtractions for
the 3+1 dimensional theory, which can then be taken to be
\begin {equation}
   \SI = \int_0^{\beta} d\tau \int d^{3-\eps}x \left[
        \psi^* \left(
        \partial_\tau - {1\over 2m} \, \nabla^2
        - \mu \right) \psi
    + \Mren^\eps \,{2\pi a\over m} \, (\psi^* \psi)^2
    \right] ,
\end {equation}
where
\begin {equation}
   \Mren \equiv {e^{\gammaE/2}\over\sqrt{4\pi}} \, \bar\Mren .
\label{eq:MSbar}
\end{equation}
[The factor of $e^{\gammaE/2}/\sqrt{4\pi}$
in (\ref{eq:MSbar}) is what distinguishes
modified minimal subtraction (\MSbar) from unmodified minimal subtraction
(MS); the difference between the two schemes amounts to nothing more
than a multiplicative redefinition of the renormalization scale.]
The three-dimensional theory (\ref{eq:S3}) is
\begin {equation}
   S_3 = \beta \int d^{3-\eps}x \left[
        \psi_0^* \left(
        - {Z_\psi \over 2m} \, \nabla^2
        - \mu_3 \right) \psi_0
    + \Mren^\eps Z_a {2\pi a\over m} \, (\psi_0^* \psi_0)^2
    + \Mren^{-\eps} f_3
    \right] ,
\end {equation}
or equivalently
\begin {equation}
   S_3 = \int d^{3-\eps}x \left[
        {1\over 2} \, |\grad\phi|^2
        + {r_\bare\over 2} \, \phi^2
        + \Mren^\eps\,{u \over 4!} \, (\phi^2)^2
    + \Mren^{-\eps} {\cal F}_3
    \right] .
\end {equation}
This theory
is super-renormalizable and requires only
a finite number of UV counter-terms.
In particular, in the \MSbar\ renormalization scheme, only the coefficient
of $\phi^2$ is explicitly renormalized, with the exact relation
\begin {equation}
   r_\bare = \rms + {1\over (4\pi)^2\eps}\left(u\over3\right)^2
\label{eq:rMSbar}
\end {equation}
between the bare coupling $r_\bare$ and the renormalized
coupling $\rms(\bar\Mren)$.

The reader may wonder why we bother with the continuum 3-dimensional theory,
since one will use a lattice-regulated 3-dimensional theory for actual
computations of non-perturbative results.  One could instead skip the
continuum 3-dimensional theory and directly match the 3+1 dimensional theory
to the particular lattice theory used for a particular simulation.
However, it is more convenient to split this matching into two steps:
(1) 3+1 dimensions to continuum 3 dimensions, and (2) continuum 3 dimensions
to lattice 3 dimensions.  The first step has the virtue of not depending
on the details of how the theory is put on the lattice.


\section {The matching calculation}
\label{sec:match}

\subsection {What we need}
\label {sec:matchA}

There are two lattice simulation results of the three-dimensional O(2)
theory (\ref{eq:O2}) that will turn out to be relevant to
our evaluation of the number density $n$ via (\ref{eq:nmatch}).
Quoting values from Ref. \cite{boselat2}, they are
\begin {mathletters}
\label {eq:lattice}
\begin {equation}
   \kappa \equiv {\Delta\phiphic \over u}
   = -0.001198(17) ,
\end {equation}
\begin {equation}
   {\cal R} \equiv {r_\c^{\msb}(\bar\Mren=u/3) \over u^2}
   = 0.001920(2) ,
\label{eq:calR}
\end {equation}
\end {mathletters}%
where
\begin {equation}
   \Delta\phiphic \equiv \left[\phiphic\right]_u - \left[\phiphic\right]_0
\end {equation}
is the difference between
the effective theory value of $\langle\phi^2\rangle$, at the critical
point, for the cases of (i) $u$ small and (ii) the ideal gas $u{=}0$.%
\footnote{
	Technically, for the non-interacting case we must take the limit
	as the critical point is approached from negative $\mu$.
	}
Unlike $\phiphic$,
the difference $\Delta\phiphic$ is an infrared quantity,
independent of how the effective theory (\ref{eq:O2}) is regularized in
the ultraviolet.
(Ref.\ \cite{prokofev} gives an independent and statistically compatible
value of $\kappa$ but did
not analyze $r_\c$.)
The reason that $\kappa$ and ${\cal R}$ are pure numbers is dimensional
analysis.  If one picks the renormalization scale to be of order $u$, then,
at the transition, the only parameter of the O(2) theory is
the dimensionful parameter $u$.  The dependence of $\Delta\phiphic$ and
$r_\c$ on $u$ is then determined by their dimensions.

In dimensional regularization in the 3 dimensional theory,
$\Delta\phiphic$ is the same as
$\phiphic$.  This is because, for the case $u{=}0$, the transition
takes place at $\mu_3=0$, and then
\begin {equation}
   \left[\phiphic\right]_0 = 2 \int {d^{3-\eps} p \over p^2} = 0
\end {equation}
in dimensional regularization.
(The last integral vanishes by dimensional analysis, since in dimensional
regularization there is no dimensionful parameter to make up the dimensions
of the integral.)
Therefore, in the formula (\ref{eq:nmatch}) for $n(T)$, we can replace the
three-dimensional
$\langle\psi_0^*\psi_0\rangle = Z_\psi^{-1} m T \langle\phi^2\rangle$
at the phase transition by
$Z_\psi^{-1} m T \kappa u$, to obtain
\begin {equation}
   \nc(T) \simeq \left[
      192\pi^3 \kappa \, {a Z_a\over \lambda^4 Z_\psi^3} \,
              {\partial\mu_3\over\partial\mu}
      - {\partial f_3\over \partial\mu}
   \right]_{\mu=\muc(T)} ,
\label {eq:ncmatch}
\end {equation}
where we have used (\ref{eq:ru}) for $u$.
To evaluate $\nc(T)$ to second order in $a \propto u$, we therefore need
\[
   Z_a,
   \quad
   Z_\psi,
   \quad\mbox{and}\quad
   \left(\partial\mu_3\over\partial\mu\right)_{\!\muc},
\]
to first order in $a$
and
\[
   \left(\partial f_3\over\partial\mu\right)_{\!\muc}
\]
to second order.
In evaluating the last quantity, we will find
that we will also want $\mu_3$ to second order in $a$, which was
computed in Ref.\ \cite{trap}.


\subsection {Matching \boldmath$Z_\psi$}

\begin {figure}
\vbox{
   \begin {center}
      \epsfig{file=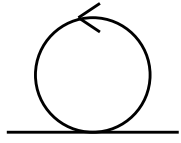,scale=.6}
   \end {center}
   \caption{
       The self energy at first order.
       \label{fig:mass1}
   }
}
\end {figure}

$Z_\psi$ can be matched at first order by matching the infrared momentum
dependence of the inverse Green function for $\psi_0$ between the
3+1 and 3 dimensional theories.
The one-loop contribution to the inverse Green function, shown in
Fig.\ \ref{fig:mass1}, is momentum independent.  So
\begin {equation}
   Z_\psi = 1 + {\cal O}(a^2) ,
\end {equation}
where ${\cal O}(a^2)$ indicates corrections that are formally second order in
perturbation theory.
A slightly more detailed discussion is given in Ref. \cite{trap}.

In this paper, we will write $O(\cdots)$ when displaying the full
parameter dependence of a correction (except possibly for logarithmic
factors) and write ${\cal O}(\cdots)$ when just showing the dependence on
a particular parameter.
So $32 a^2/\lambda^2 = O(a^2/\lambda^2) = {\cal O}(a^2)$.
In matching calculations, where we are formally doing perturbation theory
with IR regularization,
${\cal O}(a^n)$ will just mean $n$-th order in perturbation theory.

We now return to why we could
drop the gradient term
\begin {equation}
       - {\partial Z_\psi\over\partial\mu} \, {1\over2m}
                \langle |\grad \psi_0|^2 \rangle
       = - {\partial Z_\psi\over\partial\mu} \,
                {T\over 2 Z_\psi} \,
                \langle |\grad \phi|^2 \rangle
\label {eq:ngradient}
\end {equation}
from the original matching formula
(\ref{eq:nmatch0}) for $n$.
At the transition, the dimensionally regulated
$\langle |\grad\phi|^2\rangle$ must be $O(u^3)$ by dimensional
analysis.
This means that the expression (\ref{eq:ngradient}) is
at least third order in the interaction strength $a$ and so
irrelevant to our second-order calculation of $\nc$.
In fact, it is fifth order, since the dominant contribution
to $\partial Z_\psi/\partial\mu$ is ${\cal O}(a^2)$.


\subsection {Matching \boldmath$Z_a$}

\begin {figure}
\vbox{
   \begin {center}
      \epsfig{file=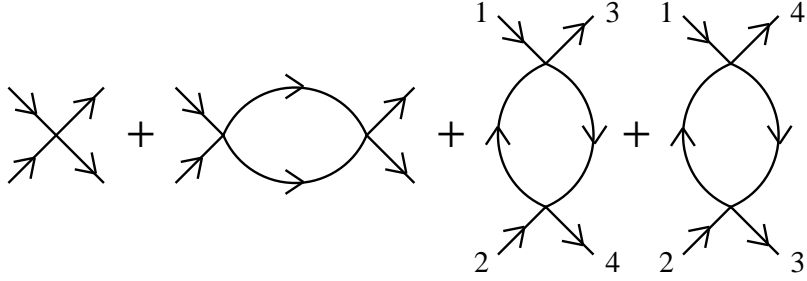,scale=.5}
   \end {center}
   \caption{
       The effective four-point interaction up to one loop.
       \label{fig:umatch}
   }
}
\end {figure}

We will determine $Z_a$ by perturbatively matching the four-point
Green function of $\psi_0$ at zero momentum.  Specifically, to match
$Z_a$ at first order, we will compute the (amputated) diagrams of
Fig.\ \ref{fig:umatch} in both the 3+1 and 3 dimensional theories.
We need to choose an IR regulator for these two computations.
For this case, we find the most convenient choice to be dimensional
regularization, which will now be used to regulate both IR and UV
infinities.

In the 3+1 dimensional theory, the diagrams of Fig.\ \ref{fig:umatch}
with zero external momenta give $-\Mren^\eps \Gamma^{(4)}$, where
\begin {equation}
   - \Gamma^{(4)} =
   - {8\pi a\over m}
   + \left(-{8\pi a\over m}\right)^2 \left[
        {1\over2} \sumint_P {1\over (ip_0+\omega_p)(-ip_0+\omega_p)}
        + 2 \sumint_P {1\over (ip_0+\omega_p)^2}
     \right]
   + {\cal O}(a^3) .
\label {eq:Zastart}
\end {equation}
We introduce the short-hand notations
\begin {eqnarray}
   \sumint_P &\equiv& T \sum_{p_0} \int_\p ,
\label {eq:defsumint}
\\
   \int_\p &\equiv& \Mren^\eps \! \int {d^dp \over (2\pi)^d} \, ,
\end {eqnarray}
where $d{=}3{-}\eps$ is the number of spatial dimensions.
The 3-dimensional theory result is the same as (\ref{eq:Zastart})
but with only zero-mode contributions
included and factors of $Z_a$ and $Z_\psi$ inserted, so that
\begin {equation}
   - \Gamma^{(4)} =
   - Z_a {8\pi a\over m}
   + \left(- Z_a {8\pi a\over m}\right)^2 \left[
        {T\over2} \int_\p {Z_\psi^{-2}\over \omega_p^2}
        + 2 T \int_\p {Z_\psi^{-2}\over \omega_p^2}
     \right]
   + {\cal O}(a^3) .
\label {eq:Zastart3}
\end {equation}
Equating the 3+1 dimensional result (\ref{eq:Zastart}) and 3 dimensional
result (\ref{eq:Zastart3}), and keeping in mind that $Z_a = 1 +{\cal O}(a)$
and $Z_\psi = 1+{\cal O}(a^2)$,
we can now solve for $Z_a$ through second order:
\begin {equation}
   Z_a = 1
   -{8\pi a\over m} \left[
        {1\over2} \sumint_P
              {1-\delta_{p_0,0}\over (ip_0+\omega_p)(-ip_0+\omega_p)}
        + 2 \sumint_P {1-\delta_{p_0,0}\over (ip_0+\omega_p)^2}
     \right]
   + {\cal O}(a^3) ,
\end {equation}
where $\delta_{i,j}$ is the Kronecker delta function.
This expression is IR convergent, as it should be.
In Appendix \ref{app:ints}, we derive the dimensionally regulated
results
\begin {equation}
   \sumint_P {1-\delta_{p_0,0}\over(ip_0+\omega_p)(-ip_0+\omega_p)} =
   \Mren^\eps \beta \lambda^{-d} \, {\zeta\!\left(\dtmo\right)\over\dtmo} ,
\label {eq:int2}
\end {equation}
\begin {equation}
   \sumint_P {1-\delta_{p_0,0}\over (ip_0+\omega_p)^2} =
   \Mren^\eps \beta \lambda^{-d} \zeta\!\left(\textstyle\dtmo\right) ,
\label {eq:intsqr}
\end {equation}
in $d=3-\eps$ spatial dimensions.
The result, after taking $d\to3$, is
\begin {equation}
   Z_a = 1
         - 12 \, \zeta(\half) \, {a\over\lambda}
         + O\left(a^2\over\lambda^2\right) .
\label {eq:Zamatch}
\end {equation}

For future reference, it's worth briefly stepping through the same calculation
if we had separately evaluated the 3+1 dimensional and 3 dimensional
contributions to the matching
(the $1$ and $\delta_{p_0,0}$ pieces of
$1-\delta_{p_0,0}$), which are individually ill-defined without
specifying a consistent IR regulator.
We will often find it convenient to use dimensional regularization to
regulate the IR (as well as the UV).  The three-dimensional integrals are
very simple in dimensional regularization,
\begin {equation}
   \int_\p {1\over\omega_p^2} \propto \int {d^d p\over p^4} = 0 ,
\label {eq:dimregzero}
\end {equation}
which follows by dimensional analysis.
The full 3+1 dimensional piece would then be the same as the
$1-\delta_{p_0,0}$ results above.  For example,
\begin {equation}
   \sumint_P {1\over (ip_0+\omega_p)^2} =
   \Mren^\eps \beta \lambda^{-d} \zeta\!\left(\textstyle\dtmo\right) .
\label {eq:intsqrfull}
\end {equation}

\begin {figure}
\vbox{
   \begin {center}
      \epsfig{file=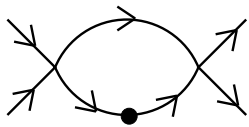,scale=.5}
   \end {center}
   \caption{
       Example of a $\mu$-dependent contribution to the inverse
       susceptibility.
       \label{fig:umatchmu}
   }
}
\end {figure}

We are now in a position to explain why we could drop the quartic
term
\begin {equation}
       - {\partial Z_a\over\partial\mu} \, {2\pi a\over m} \,
                \langle (\psi_0^* \psi_0)^2 \rangle
\label {eq:nquartic}
\end {equation}
from the original matching formula
(\ref{eq:nmatch0}) for $n$.
{}From (\ref{eq:Zamatch}), or the diagrams of Fig.\ \ref{fig:umatch}, we
see that $Z_a$ is $\mu$-independent at first order.
The leading $\mu$-dependent contribution comes from graphs such as
Fig.\ \ref{fig:umatchmu}, which will produce an ${\cal O}(\mu a)$
contribution to $Z_a$ and so an ${\cal O}(a)$ contribution to
$\partial Z_a/\partial\mu$.  There is an explicit $a$ in
(\ref{eq:nquartic}), which brings us up to ${\cal O}(a^2)$.
Finally, there is the factor of
\begin {equation}
   \langle (\psi_0^*\psi_0)^2 \rangle
   = Z_\psi^{-2} m^2 T^2 \langle \phi^4 \rangle .
\end {equation}
At the transition, the dimensionally-regulated result for
$\langle\phi^4\rangle$ must be order $u^2$ by dimensional
analysis.  This means that the contribution (\ref{eq:nquartic})
to $n$ is fourth-order in the interaction strength and so irrelevant
to a second-order calculation of $\nc(T)$.
This argument is almost identical to a similar argument given in
Ref.\ \cite{boselat2} for the matching of the 3 dimensional continuum theory
to a 3 dimensional lattice theory.

\subsection {Matching the \boldmath$\mu$ dependence of \boldmath$\mu_3$}

\begin {figure}
\vbox{
   \begin {center}
      \epsfig{file=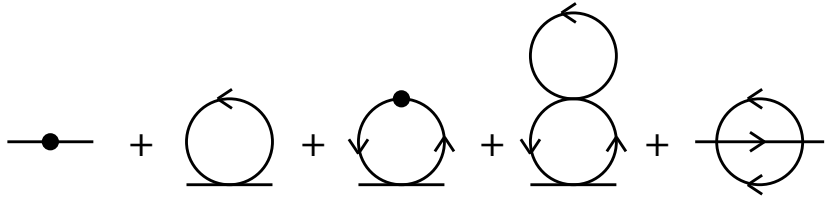,scale=.6}
   \end {center}
   \caption{
       The inverse susceptibility through two-loop order.
       \label{fig:mu3match}
   }
}
\end {figure}

The matching of $\mu_3$ can be accomplished by computing the inverse
susceptibility $\chi^{-1}=\Gamma^{(2)}$
in both theories.  At second order in $a$, this corresponds to
the diagrams of Fig.\ \ref{fig:mu3match}.
For the purpose of counting orders of $a$, we treat the chemical potential
$\mu$ as ${\cal O}(a)$.  That's because we are ultimately interested in
using the 3 dimensional effective
theory at the phase transition -- that is, for
$\mu = \muc$.  The ideal gas result is $\muc=0$, and the effect of
interactions is that $\muc = {\cal O}(a)$.

We discussed earlier that we need $\partial\mu_3/\partial\mu$ to
first order.  For this computation, we will find we can ignore the
$\mu$-independent diagrams of Fig.\ \ref{fig:mu3match}.
Specifically, Fig.\ \ref{fig:mu3match} gives
\begin {equation}
   -\Gamma^{(2)} =
   \mu - {8\pi a\over m} \,\mu \sumint_P {1\over (ip_0+\omega_p)^2}
       + \mbox{($\mu$-independent)} + {\cal O}(a^3)
\end {equation}
for the 3+1 dimensional theory and
\begin {equation}
   -\Gamma^{(2)} =
   \mu_3 - Z_a {8\pi a\over m} \,\mu_3 T \int_\p
                 {Z_\psi^{-2}\over \omega_p^2}
       + \mbox{($\mu_3$-independent)} + {\cal O}(a^3)
\end {equation}
for the 3 dimensional theory.
Equating the two results, order by order in $a$, yields
\begin {equation}
  \mu_3 = Z_\mu \mu + \mbox{($\mu$-independent)} + {\cal O}(a^3) ,
\end {equation}
with
\begin {equation}
  Z_\mu = 
      1
      - {8\pi a\over m} \sumint_P{1-\delta_{p_0,0}\over(ip_0+\omega_p)^2} .
\label {eq:Zmu1}
\end {equation}
We then have
\begin {equation}
   {\partial\mu_3\over\partial\mu} = Z_\mu + {\cal O}(a^3) .
\end {equation}
The integral in (\ref{eq:Zmu1}) is the same as one of those encountered
matching $Z_a$, and the result is
\begin {equation}
   {\partial\mu_3\over\partial\mu} =
      1 - 4 \, \zeta(\half) \, {a\over\lambda}
      + O\left(a^2\over\lambda^2\right) .
\end {equation}

A full result for $\mu_3$ at exactly $\mu=\muc$ was derived in
Ref.\ \cite{trap}.  We will rederive the crux of that result in
section \ref{sec:sunset},
since we will need it in a different analytic form than
presented in the original derivation.


\subsection {Matching the \boldmath$\mu$ dependence of \boldmath$f_3$}

\begin {figure}
\vbox{
   \begin {center}
      \epsfig{file=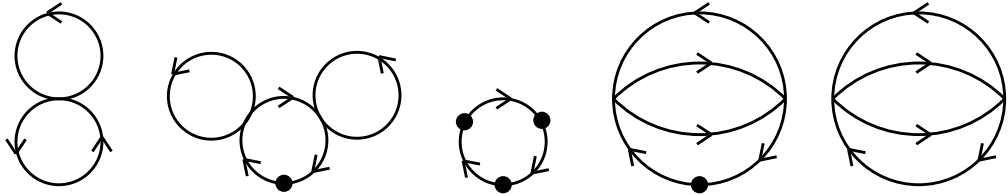,scale=.6}
   \end {center}
   \caption{
       Examples of diagrams which contribute to the free energy at
       up to third order.
       \label{fig:f3sample}
   }
}
\end {figure}

\begin {figure}
\vbox{
   \begin {center}
      \epsfig{file=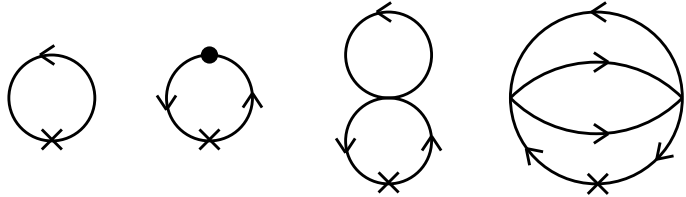,scale=.6}
   \end {center}
   \caption{
       Diagrams contributing to the $\mu$-derivative of the free energy,
       through second order at the transition.
       Not shown is the tree-level contribution,
       $\partial f_3/\partial\mu$ for the 3-dimensional theory.
       \label{fig:f3main}
   }
}
\end {figure}

\begin {figure}
\vbox{
   \begin {center}
      \epsfig{file=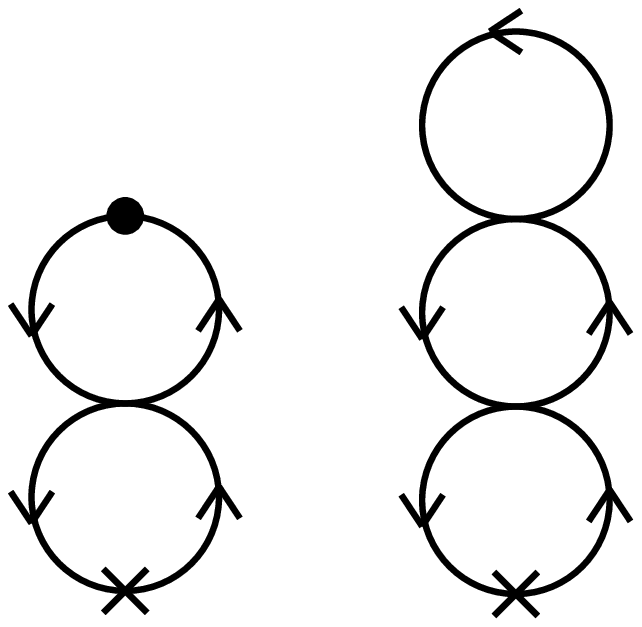,scale=.6}
   \end {center}
   \caption{
       Contributions to the $\mu$-derivative of the free energy
       which cancel at second order {\it at the transition}.
       \label{fig:f3cancel1}
   }
}
\end {figure}

\begin {figure}
\vbox{
   \begin {center}
      \epsfig{file=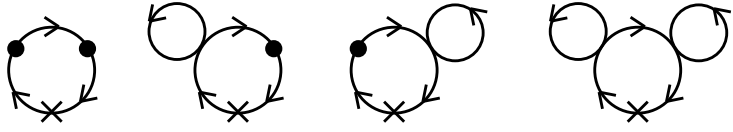,scale=.6}
   \end {center}
   \caption{
       The remaining second-order
       contributions to the $\mu$-derivative of the free energy,
       which also cancel at second order {\it at the transition}.
       \label{fig:f3cancel2}
   }
}
\end {figure}

We now come to the more intricate part of our calculation, which is the
calculation of $\partial f_3/\partial \mu$ at the transition.
We can match $f_3$ by computing the free energy density $f$ in both theories.
Some examples of diagrams which contribute to the $\mu$-dependence of
the free energy are shown in Fig.\ \ref{fig:f3sample}.
We will find it convenient to instead compute the derivative directly, a
diagrammatic representation of which is given by
Figs.\ \ref{fig:f3main}--\ref{fig:f3cancel2}, which contain all diagrams
contributing through ${\cal O}(a^2)$ in perturbation theory.
The filled circles are still associated with a factor of $\mu$, but
the crosses, which represent factors of $\mu$ that have been hit by
a derivative, are associated with a factor of $1$.
One can also think of these diagrams as representing mixing of the
operator $\phi^2$ (represented by the crosses) with the unit operator,
which were the words used to describe analogous calculations in
Ref.\ \cite{boselat2}.

At the transition, the diagrams of Fig.\ \ref{fig:f3cancel1}
cancel at second order in $a$, as do the diagrams of Fig.\ \ref{fig:f3cancel2},
because of the first-order relation of Fig.\ \ref{fig:chi1}.
So we may simplify our task by focusing on just the diagrams of
Fig.\ \ref{fig:f3main}.

At this many loops, organizing all the terms of the matching calculation
becomes tedious unless one from the start uses dimensional regularization for
the IR (as well as the UV).  In the 3-dimensional theory, the loop integrals
in Fig.\ \ref{fig:f3main} then all vanish by dimensional analysis, leaving
us with the formal result
\begin {equation}
   -{\partial f\over\partial\mu} = -{\partial f_3\over\partial\mu}
\end {equation}
for dimensionally regulated perturbation theory.
In the 3+1 dimensional theory, we have
\begin {equation}
   -\left({\partial f\over\partial\mu}\right)_{\muc} =
   \sumint_P {1\over(ip_0+\omega_p)}
   + (\muc-\mucone) \sumint_P {1\over(ip_0+\omega_p)^2}
   + \dnbball
   + {\cal O}(a^3)
\label {eq:f3start}
\end {equation}
at the transition, where
\begin {equation}
  \mucone = {8\pi a\over m} \sumint_P {1\over(i p_0 + \omega_p)}
\end {equation}
is the first-order result for $\muc$, determined by Fig.\ \ref{fig:chi1},
and
\begin {equation}
   \dnbball \equiv 
   {1\over2}\left(-{8\pi a\over m}\right)^2
       \sumint_{PQK} {1\over (i p_0 + \omega_p)^2 (i q_0 + \omega_q)
               (i k_0 + \omega_k) (i(p_0+q_0+k_0)+\omega_{\p+\q+\k})}
\label {eq:bballstart}
\end {equation}
is the contribution of the last diagram of Fig.\ \ref{fig:f3main},
which we refer to as the basketball diagram.%
\footnote{
   Historical note: The origin of this terminology in the literature
   is associated with the physical appearance of old American Basketball
   Association
   basketballs, not current National Basketball Association ones.
}
In section \ref{sec:bball},
we will show how to evaluate this integral in dimensional
regularization, with the result that
\begin {equation}
   \dnbball =
   {32\pi a^2\over\lambda^5} \left[
       {\sqrt\pi\over2}
       - K_2
       + (2 K_1+\ln 2) \, \zeta(\half)
       - {\ln 2\over 2\sqrt\pi} \, \left[\zeta(\half)\right]^2
   \right] + {\cal O}(\eps) ,
\label {eq:bballfinal}
\end {equation}
where $K_1$ and $K_2$ are numerical constants defined by%
\footnote{
   For efficient numerical evaluation of $K_1$, it's useful to make a change
   of integration variables in (\ref{eq:K1}), such as $u = \sqrt{1-t}$,
   which makes the integrand analytic at both endpoints.
}
\begin {eqnarray}
   K_1 &\equiv& {1\over4\pi}
   \int_0^1{dt\over t} \>
       \left[ \left[\Li_{1/2}(t)\right]^2 - {\pi t\over 1-t} \right]
   \simeq  -0.630~568~207~496~069  
   ,
\label {eq:K1}
\\[8pt]
   K_2 &\equiv& 
       {1\over4\pi} \int_0^1 {ds\over s} \> {dt\over t} \biggl\{
             \Li_{1/2}(s)\,\Li_{1/2}(t)\,\Li_{-1/2}(st)
\nonumber\\ && \hspace{4em}
             - {s t\sqrt\pi\over 2(1-st)^{3/2}}
               \left[ \sqrt{\pi\over 1-s} + \zeta(\half) \right]
               \left[ \sqrt{\pi\over 1-t} + \zeta(\half) \right]
         \biggr\}
\nonumber\\
   &\simeq& -0.135~083~353~73
   .
\label {eq:K2}
\end {eqnarray}
$\Li_\nu(z)$ is the polylogarithm function defined by
\begin {equation}
   \Li_\nu(z) = \sum_{n=1}^\infty {z^n\over n^\nu} .
\end {equation}
[The polylogarithm function is often called $g_\nu(z)$
in statistical mechanics.]

The other integrals in the matching (\ref{eq:f3start}) are given by
(\ref{eq:intsqrfull}) and
\begin {equation}
   \sumint_P{1\over i p_0+\omega_p}
   = \Mren^\eps \lambda^{-d} \, \zeta\left({\textstyle{d\over2}}\right) ,
\label {eq:basicint}
\end {equation}
which is discussed in Appendix \ref{app:ints}.
Putting it all together and taking $d\to3$,
\begin {equation}
   -\left({\partial f_3\over\partial\mu}\right)_{\muc} =
   \zeta(\threehalf)\,\lambda^{-3}
   + \zeta(\half)\,\lambda^{-3} \, (\beta\muc-\beta\mucone)
   + \dnbball
   + {\cal O}(a^3) \, .
\end {equation}
In Ref. \cite{trap}, a second-order matching calculation was carried out
for $\mu_3$ that determined $\muc$ in terms of the critical value of $r$
in the 3 dimensional theory.  The result was
\begin {equation}
   \beta\muc =
   \beta\mucone
   + {32\pi a^2\over\lambda^2} \left[
      \ln(\bar\Mren \lambda) + C_1 - 72\pi^2 \,{r_\c(\bar\Mren)\over u^2}
     \right] ,
\label {eq:muc}
\end {equation}
with
$C_1$ a numerical constant that was given in that reference
in terms of a somewhat inelegant double integral.
In section \ref{sec:sunset}, we show that $C_1$ can also be expressed as
\begin {equation}
   C_1 = \half - \half \ln(32\pi) - 2 K_1 .
\label {eq:C1}
\end {equation}
Combining the last several formulas, and choosing $\bar\Mren=u/3$ to make
contact with the quoted lattice measurement (\ref{eq:calR}) of $r_\c$, we have
\begin {eqnarray}
   -\left({\partial f_3\over\partial\mu}\right)_{\muc} &=&
   \zeta(\threehalf)\,\lambda^{-3}
   + {32\pi a^2\over\lambda^5} \biggl\{
      \left[
         \ln\left(a \over \lambda\right) 
         + \half\ln(128\pi^3) + \half - 72\pi^2 {\cal R}
      \right] \zeta(\half)
\nonumber\\ && \hspace{8em}
      + {\sqrt\pi\over2}
      - K_2
      - {\ln 2\over2\sqrt\pi}\left[\zeta(\half)\right]^2
   \biggr\}
   + O\left(a^3\over\lambda^6\right)
   .
\label {eq:f3result}
\end {eqnarray}


\section {Final Results}
\label {sec:final}

We now have all the elements we need for $\nc(T)$ as determined by
(\ref{eq:ncmatch}).  The result is
\begin {eqnarray}
   \nc(T) = \lambda^{-3} \left\{
      b_0
      + b_1 a\lambda^{-1}
      + [b_2' \ln(a\lambda^{-1}) + b_2''] a^2\lambda^{-2}
      + O(a^3\lambda^{-3})
      \right\} ,
\label {eq:ncexpand}
\end {eqnarray}
with
\begin {mathletters}
\label {eq:bi}
\begin {eqnarray}
   b_0 &=& \zeta(\threehalf) ,
\\
   b_1 &=& 192\pi^3\kappa ,
\\
   b_2' &=& 32\pi\,\zeta(\half) ,
\\
   b_2'' &=&
   32 \pi \biggl\{
      \left[
               \half\ln(128\pi^3) + \half - 72\pi^2 {\cal R}
               -96\pi^2\kappa
      \right] \zeta(\half)
      + {\sqrt\pi\over2}
      - K_2
      - {\ln 2\over2\sqrt\pi}\left[\zeta(\half)\right]^2
   \biggr\}
   .
\end {eqnarray}
\end {mathletters}%
Inverting this formula,
\begin {equation}
   \Tc(n) = T_0(n) \, \bigg\{ \,
      1
      + c_1 an^{1/3}
      + \left[c_2' \ln(an^{1/3}) + c_2'' \right] a^2n^{2/3}
      + O(a^3 n)
   \bigg\} \, ,
\end {equation}
with
\begin {eqnarray}
   c_1 &=& -{\textstyle{2\over3}}\, b_0^{-4/3} b_1
   \,,
\\
   c_2' &=& -{\textstyle{2\over3}}\, b_0^{-5/3} b_2'
   \,,
\\
   c_2'' &=&
      -{\textstyle{2\over3}}\, b_0^{-5/3} b_2'' 
      +{\textstyle{7\over9}}\, b_0^{-8/3} b_1^2 
      -{\textstyle{1\over3}}\, c_2'\ln b_0
   \, ,
\end {eqnarray}
where the $\ln(b_0)$ term arises because we have changed the argument of
the log from $\ln(a/\lambda)$ to $\ln(an^{1/3})$.
Putting in the lattice results of (\ref{eq:lattice}), we get the numerical
values
\begin {equation}
   b_1 = -7.1(1) ,
   \qquad
   b_2' = -146.8108\cdots ,
   \qquad
   b_2'' = -587(2) ,
\end {equation}
\begin {equation}
   c_1 = 1.32(2) ,
   \qquad
   c_2' = 19.7518\cdots ,
   \qquad
   c_2'' = 75.7(4) \, .
\end {equation}
These are our final results.

In our discussion of effective theories, we left out a variety of
corrections, such as 3-body interactions or cross-section energy dependence
in the original 3+1 dimensional theory, or $\phi^6$ and
higher-dimensional operators in the 3-dimensional effective theory.
Ref.\ \cite{trap} contains a detailed discussion of the parametric size
of the resulting corrections to $\Tc(\Np)$ for an arbitrarily wide
harmonic trap.  The same analysis holds for $\Tc(n)$, with the result that
there are no corrections at second order.  However, a third-order result
would depend not only on the scattering length $a$ but also on the
effective range of the two-body scattering potential.

The relative size of the second-order result obviously depends on the
diluteness of the gas and the value of the scattering length, which will
vary from experiment to experiment.  However, just for fun, let us
produce numbers for one particular case.
In 1996, Ensher \etal\ \cite{ensher} studied the BEC transition for
for dilute gases of ${}^{87}$Rb atoms in the
$F{=}2$ hyperfine state, trapped in a harmonic trap.
The relevant scattering length is $a = (103 \pm 5) \, a_0$
\cite {julienne},
where $a_0 = 0.0529177$ nm is the Bohr radius.
Their transition was at $T \simeq 280$ nK.
These parameters correspond to $a/\lambda_0 \simeq 0.015$.
Let's now consider a theoretical prediction of the central density $n$
(which, sadly, is not directly accessible experimentally).
Eq. (\ref{eq:ncexpand}) gives a first-order correction to the ideal
gas result $n_0(T)$ of roughly
$-4.2$\% and a second-order correction of roughly $+0.2$\%.
As discussed in Ref.\ \cite{trap}, where a similar analysis is made
of $\Tc(\Np)$, this particular trap may not actually be wide enough for
our second-order result to be valid.


\section {Evaluating \boldmath$\dnbball$ with dimensional regulation}
\label {sec:bball}

\begin {figure}
\vbox{
   \begin {center}
      \epsfig{file=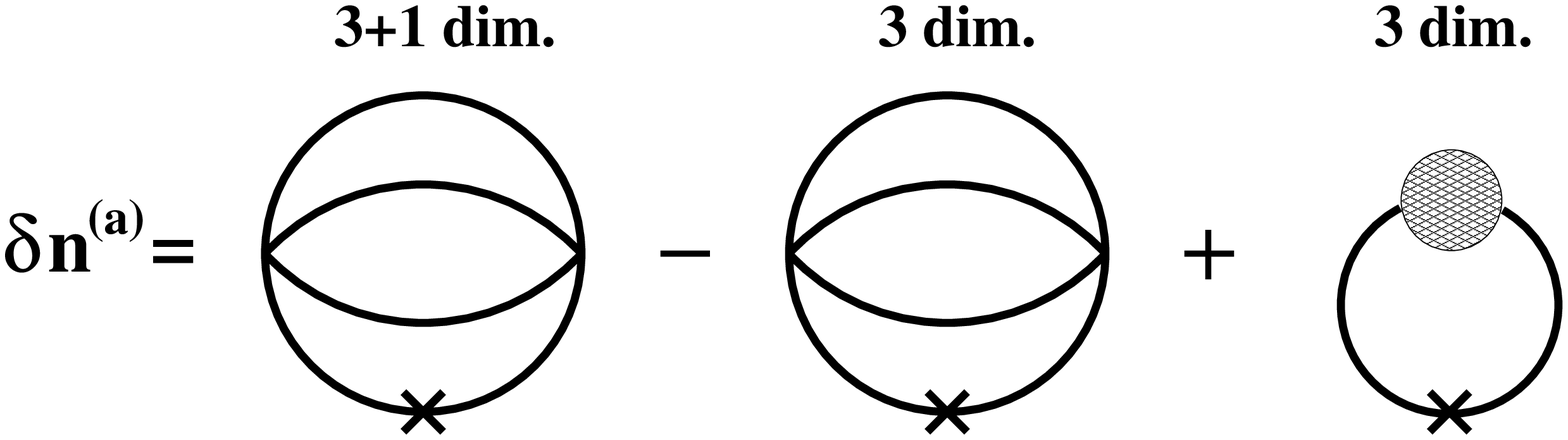,scale=.4}
   \end {center}
   \caption{
       IR convergent combination of diagrams corresponding to
       $\dnbball$.
       \label{fig:IRsafe}
   }
}
\end {figure}

\begin {figure}
\vbox{
   \begin {center}
      \epsfig{file=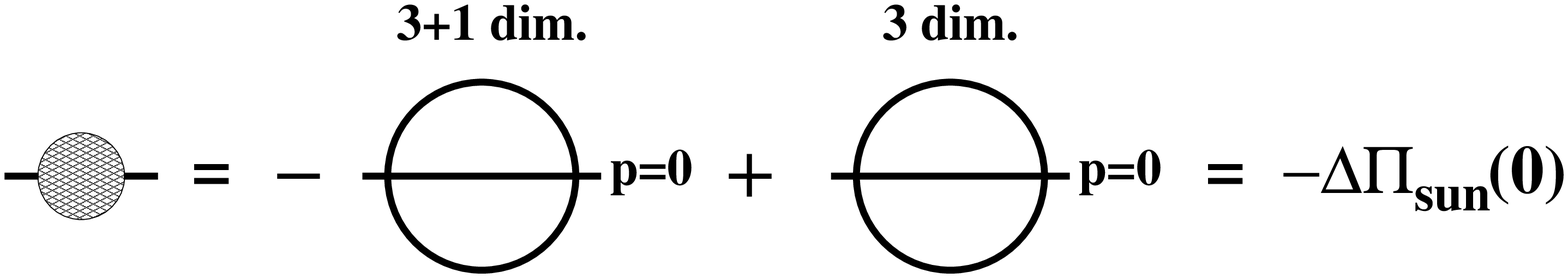,scale=.4}
   \end {center}
   \caption{
       A piece of the second-order correction to the relationship between
       $\mu$ and $r$, which is to be understood to be evaluated at zero
       external momentum.
       \label{fig:IRsafemu}
   }
}
\end {figure}

We will now explain how to obtain the result (\ref{eq:bballfinal}) for
the basketball diagram in dimensional regularization.
We do not know how to carry out the integrations for
the basketball (\ref{eq:bballstart}) in arbitrary dimension $d$.
Instead, we will find it convenient to manipulate the integrals into
a form where we can dispense with dimensional regularization and do
integrations in three dimensions.  We are currently relying on dimensional
regularization to regulate the IR and UV divergences of our calculation.
Our first step
on the road to dispensing with dimensional regularization
of the basketball diagram
will be to convert the IR role of dimensional regularization to mass
regularization, which will turn out to be more convenient for this
particular calculation.
To make this conversion consistently, we will first relate the
basketball diagram to a combination of diagrams that is IR
convergent.
In particular, consider the combination of diagrams $\dnbball'$
depicted in Figs.\ \ref{fig:IRsafe} and \ref{fig:IRsafemu}
(ignoring all $Z_\psi$ and $Z_a$ factors).
In equations, this is
\begin {eqnarray}
   \dnbball' &\equiv&
   {1\over2}\left(-{8\pi a\over m}\right)^2 \biggl[
       \sumint_{PQK} {1\over (i p_0 + \omega_p)^2 (i q_0 + \omega_q)
               (i k_0 + \omega_k) (i(p_0+q_0+k_0)+\omega_{\p+\q+\k})}
\nonumber\\ && \hspace{10em}
       - T^3 \int_{\p\q\k} {1\over \omega_p^2 \omega_q \omega_k
               \omega_{\p+\q+\k}}
   \biggr]
   - \Delta\Pi_\sunset(0)\, T \int_\p {1\over \omega_p^2}
   ,
\label {eq:IRsafe}
\end {eqnarray}
where
\begin {eqnarray}
   \Delta\Pi_\sunset(0) &=&
    {1\over2} \left(-{8\pi a\over m}\right)^2 \biggl[
      \sumint_{QK} {1\over (i q_0 + \omega_q)
               (i k_0 + \omega_k) (i(q_0+k_0)+\omega_{\q+\k})}
\nonumber\\ && \hspace{8em}
       - T^2 \int_{\q\k} {1\over \omega_q \omega_k \omega_{\q+\k}}
   \biggr]
\nonumber\\&&
\label{eq:IRsafemu}
\end {eqnarray}
is the difference between the ``sunset'' diagram contributions to
the self-energy at zero momentum in the 3+1 dimensional and 3 dimensional
theories.
The above can be combined into
\begin {eqnarray}
   \dnbball' &=&
   {1\over2}\left(-{8\pi a\over m}\right)^2
       \sumint_{PQK}
       {1-\delta_{p_0,0}\delta_{q_0,0}\delta_{k_0,0}
          \over(i p_0+\omega_p)^2 (i q_0 + \omega_q) (i k_0 + \omega_k)}
\nonumber\\ && \hspace{7em} \times
       \biggl[
          {1 \over i(p_0+q_0+k_0)+\omega_{\p+\q+\k}}
          - {\delta_{p_0,0} \over i(q_0+k_0)+\omega_{\q+\k}}
       \biggr] .
\label{eq:IRsafeB}
\end {eqnarray}
This expression is IR convergent.
(Actually, it is absolutely convergent in the IR only if one first
averages the integrand over $\p \to -\p$.  This is a technical point
that won't have any impact, and we shall implicitly assume such averaging
wherever required.)

In fact, in dimensional regularization, the original basketball
(\ref{eq:bballstart}) is equal to the IR-convergent version
(\ref{eq:IRsafe}).  That's because the explicit three-dimensional
integrals of the terms that have been added to $\dnbball$ in
(\ref{eq:IRsafe}) all vanish in dimensional regularization
by dimensional analysis, as in (\ref{eq:dimregzero}).

So we can now focus on (\ref{eq:IRsafeB}).
Because it is IR convergent, it will not be changed if we add an
infinitesimal negative chemical potential $\infmu$ to the
propagators.  We then re-expand the terms into the form
of (\ref{eq:IRsafe}), where the IR divergences of the individual
terms are now regulated by $\infmu$:
\begin {equation}
   \dnbball =
   \lim_{\infmu\to0^-}
     \Bigl[\mbox{Eq.\ (\ref{eq:IRsafe}) with $\omega\to\tomega$ and
          $\Pi\to\tPi$}\Bigr] ,
\label {eq:IRsafe2}
\end {equation}
\begin {equation}
   \Delta\tPi_\sunset(0) \equiv
     \Bigl[\mbox{Eq.\ (\ref{eq:IRsafemu}) with $\omega\to\omega^+$}\Bigr],
\label {eq:IRsafemu2}
\end {equation}
where
\begin {equation}
   \tomega \equiv \omega - \infmu .
\end {equation}
Henceforth, the limit $\delta\mu \to 0^-$ will be implicit when we
discuss $\dnbball$.


\subsection {The mass-regulated basketball}
\label{sec:dn1}

We will now focus on the first term of (\ref{eq:IRsafe2}),
\begin {equation}
   \dn_1 \equiv 
   {1\over2}\left(-{8\pi a\over m}\right)^2
       \sumint_{PQK} {1\over (i p_0 + \tomega_p)^2 (i q_0 + \tomega_q)
               (i k_0 + \tomega_k) (i(p_0+q_0+k_0)+\tomega_{\p+\q+\k})} ,
\end {equation}
which is just
our original basketball diagram with an infinitesimal chemical potential
$\delta\mu$,
and with the UV still regulated with dimensional regularization.
This can be expressed as the derivative with respect to $\delta\mu$
of the corresponding contribution
\begin {equation}
   P_1 \equiv
   {1\over8}\left(-{8\pi a\over m}\right)^2
       \sumint_{PQK} {1\over (i p_0 + \tomega_p) (i q_0 + \tomega_q)
               (i k_0 + \tomega_k) (i(p_0+q_0+k_0)+\tomega_{\p+\q+\k})}
\end {equation}
to the pressure, corresponding to the last diagram of
Fig.\ \ref{fig:f3sample}.
For arbitrary chemical potential (not just the infinitesimal case),
this contribution was derived in Ref.\ \cite{trap} for UV dimensional
regularization, and simply reproduces a corresponding
portion of a 1957 calculation
by Huang, Yang, and Luttinger \cite{huang},
which used a different UV regulator.
The result (after taking $d \to 3$) is
\begin {equation}
   P_1 = {8 T a^2\over \lambda^5}
            \sum_{i=1}^\infty \sum_{j=1}^\infty \sum_{k=1}^\infty
                  {z^{i+j+k}\over (i+k)(j+k)(ijk)^{1/2}}
   ,
\label {eq:P}
\end {equation}
where $z$ is the corresponding fugacity.  In our case,
\begin {equation}
   z = e^{\beta\,\infmu} ,
\end {equation}
which is infinitesimally less than one and is serving the role of an
IR regulator.
Differentiation gives
\begin {equation}
   \dn_1 = {8 a^2\over \lambda^5} S(z) ,
\label {eq:na}
\end {equation}
with
\begin {equation}
   S(z) \equiv \sum_{ijk} {(i+j+k)z^{i+j+k} \over (i+k)(j+k)(ijk)^{1/2}} .
\end {equation}
Our task is to find the $\infmu\to 0^-$ ($z\to1^-$) behavior of this sum,
extracting any divergences and the finite remainder.
It will be convenient to define
\begin {equation}
   \alpha \equiv -\beta\,\infmu,
   \qquad
    z = e^{-\alpha} ,
\end {equation}
so that the limit of interest is $\alpha \to 0^+$.

Now rewrite the sum as
\begin {eqnarray}
   S(z)
   &=& \sum_{ijk}
       {(i+j+k) z^{i+j+k}\over (ijk)^{1/2}}
       \int_0^1 {ds\over s} \> {dt\over t} \>
       s^{i+k} t^{j+k}
\nonumber\\
   &=& 
       \int_0^1 {ds\over s} \> {dt\over t}
       \sum_{ijk}
       s^{i+k} t^{j+k} z^{i+j+k}
       \left[
           2 \left(i\over jk\right)^{1/2} + \left(k\over ij\right)^{1/2}
       \right] .
\end {eqnarray}
If one does the sums before the integrals, the sums can then be factorized,
giving
\begin {equation}
   S(z)
   = \int_0^1 {ds\over s} \> {dt\over t} \>
       \left[ 2\, \Li_{-1/2}(sz) \, \Li_{ 1/2}(tz) \, \Li_{ 1/2}(stz)
                + \Li_{ 1/2}(sz) \, \Li_{ 1/2}(tz) \, \Li_{-1/2}(stz) \right] .
\label {eq:Sa}
\end {equation}

If we removed the mass regularization by simply setting $z=1$, the above
integral would have singularities associated both with (i) $s \to 1$
with $t$ fixed, and (ii) $s$ and $t$ approaching 1 simultaneously.
The first type of divergence arises only from the first term in
the regulated integral (\ref{eq:Sa}) and can be eliminated by integrating
this term by parts.  Using
\begin {equation}
   {d\over ds} \, \Li_\nu(sz) = {1\over s} \, \Li_{\nu-1}(sz) ,
\end {equation}
integration by parts gives
\begin {equation}
   \int_0^1 {ds\over s} \> \Li_{-1/2}(sz) \, \Li_{1/2}(stz)
   = \Li_{1/2}(z) \, \Li_{1/2}(tz)
       - \int_0^1 {ds\over s} \> \Li_{1/2}(sz) \, \Li_{-1/2}(stz) .
\end {equation}
Application to (\ref{eq:Sa}) then yields
\begin {equation}
   S(z)
   = S_1(z)
   - \int_0^1 {ds\over s} \> {dt\over t} \>
            \Li_{1/2}(sz) \, \Li_{1/2}(tz) \, \Li_{-1/2}(stz) ,
\label {eq:SaA}
\end {equation}
where
\begin {equation}
   S_1(z) \equiv 2 \, \Li_{1/2}(z) \int_0^1 {dt\over t} \>
         \left[\Li_{1/2}(tz)\right]^2 .
\label {eq:S1}
\end {equation}
To analyze the behavior of this expression as $z\to1$, it will be useful
to have the following series expansion of polylogarithms \cite{robinson},
\begin {equation}
   \Li_\nu(e^{-\alpha})
       = \Gamma(1-\nu)\,\alpha^{\nu-1}
       + \sum_{n=0}^\infty {(-)^n\over n!}\,\zeta(\nu-n)\,\alpha^n .
\end {equation}
The relevant special cases are
\begin {eqnarray}
   \Li_{+1/2}(z) &=& \sqrt{\pi\over 1-z}  + \zeta(\half) + O(\sqrt{1-z})
                = \sqrt{\pi\over\alpha} + \zeta(\half) + O(\sqrt{\alpha}) ,
\\
   \Li_{-1/2}(z) &=& {\sqrt{\pi}\over 2(1-z)^{3/2}}
                         + O\blparen(1-z)^{-1/2}\brparen
                = {\sqrt{\pi}\over2 \alpha^{3/2}} + O(\alpha^{-1/2}) .
\end {eqnarray}

Let's separate out the regularization
dependence of (\ref{eq:S1}) for $S_1$ by writing
\begin {eqnarray}
   \int_0^1 {dt\over t} \left[\Li_{1/2}(tz)\right]^2
   &=&  \int_0^1 {\pi \> dt\over 1 - z t}
      + \int_0^1 {dt\over t} \left\{ \left[\Li_{1/2}(tz)\right]^2
               - {\pi t\over 1-zt} \right\} .
\nonumber\\
   &=&  -{\pi\over z} \, \ln(1-z)
      + \int_0^1 {dt\over t} \left\{ \left[\Li_{1/2}(tz)\right]^2
               - {\pi t\over 1-zt} \right\} .
\label {eq:S1bit}
\end {eqnarray}
If we just wanted an expression through $O((1-z)^0) = O(\alpha^0)$, we
could now set $z = 1$ in the very last integral.  However, the $\Li_{1/2}(z)$
factor in (\ref{eq:S1}) has an $O(\alpha^{-1/2})$
singularity, which means we need the expansion of (\ref{eq:S1bit})
through $O(\alpha^{1/2})$.  This term is easily obtained by differentiating
the last integral in (\ref{eq:S1bit})
with respect to $z$ and then analyzing the dominant piece
of the result, which is a singularity (cut off by $z$) as $t \to 1$.
The result is
\begin {equation}
   \int_0^1 {dt\over t} \left[\Li_{1/2}(tz)\right]^2
   = - \pi \ln\alpha
      + 4\pi K_1
      - 4\,\zeta(\half)\,\sqrt{\pi\alpha}
      + O(\alpha) ,
\end {equation}
where $K_1$ is as defined in (\ref{eq:K1}).
So
\begin {equation}
   S_1 =  \left[ \sqrt{\pi\over \alpha} + \zeta(\half) \right]
                   ( - 2\pi \ln \alpha + 8\pi K_1 )
              - 8\pi \zeta(\half)
              + O(\sqrt{\alpha}).
\label {eq:S1final}
\end {equation}

Now let's return to the expression (\ref{eq:SaA}) for the sum $S$
and work on isolating the remaining (regulated) divergences associated with
$s$ and $t$ simultaneously approaching 1.  We isolate the singular pieces
of the integrand in (\ref{eq:SaA}) by writing
\begin {equation}
   S(z) = S_1(z) + S_2(z) - 4\pi K_2
       + O(\sqrt{\alpha}) ,
\label {eq:Safinal}
\end {equation}
\begin {equation}
   S_2(z) \equiv
       - \int_0^1 ds \> dt \>
             {\sqrt\pi\over 2(1-stz)^{3/2}}
               \left[ \sqrt{\pi\over 1-sz} + \zeta(\half) \right]
               \left[ \sqrt{\pi\over 1-tz} + \zeta(\half) \right]
        ,
\end {equation}
with $K_2$ defined as in (\ref{eq:K2}).
[The $\zeta(\half)^2$ term in $S_2$
doesn't actually give a singular piece, but
including it makes the remainder (\ref{eq:K2}) a little more compact to write.]
Explicit integration gives
\begin {equation}
   S_2 = - {\pi^{3/2} \ln 4\over\sqrt\alpha}
             + 2\pi\,\zeta(\half)\,\ln(4\alpha)
             - \sqrt\pi \left[\zeta(\half)\right]^2 \, \ln 4
             + 2 \pi^{3/2}
             + O(\sqrt\alpha) ,
\label {eq:S2final}
\end {equation}
and so
\begin {eqnarray}
   S_1 + S_2 &=& 
      4\pi \biggl\{
          \sqrt{\pi\over \alpha} \left[
              - \half \ln (2\alpha) + 2 K_1
            \right]
\nonumber\\ && \hspace{4em}
          + {\sqrt\pi\over2}
          + (2 K_1 - 2 + \ln 2) \, \zeta(\half) 
          - {\ln2\over 2\sqrt\pi} \left[\zeta(\half)\right]^2
      \biggr\}
      + O(\sqrt{\alpha}).
\label {eq:S12final}
\end {eqnarray}

The combination of (\ref{eq:na}), (\ref{eq:Safinal}) and
(\ref{eq:S12final}) then gives our result for the mass-regulated
3+1 dimensional basketball,
\begin {eqnarray}
   \dn_1 &=&
   {32\pi a^2\over\lambda^5} \biggl\{
      \sqrt{\pi\over \alpha} \left[
              - \half \ln (2\alpha) + 2 K_1
            \right]
\nonumber\\ && \hspace{5em}
       + {\sqrt\pi\over2}
       - K_2
       + (2 K_1-2+\ln 2) \, \zeta(\half)
       - {\ln 2\over 2\sqrt\pi} \, \left[\zeta(\half)\right]^2
   \biggr\}
   + O(\sqrt\alpha) ,
\label {eq:dn1final}
\end {eqnarray}
where $K_2$ is defined as in (\ref{eq:K2}).


\subsection {The remaining pieces}
\label {sec:remaining}

The second diagram of Fig.\ \ref{fig:IRsafe}, when mass regulated, gives a
contribution to $\dnbball$ of
\begin {eqnarray}
   \dn_2 &\equiv&
   - {1\over2}\left(-{8\pi a\over m}\right)^2
       T^3 \int_{\p\q\k} {1\over (\tomega_p)^2 \tomega_q \tomega_k
               \tomega_{\p+\q+\k}}
\nonumber\\
   &=& -2^{10} \pi^2 a^2 m^3 T^3 \int_{\p\q\k}
       {1\over (p^2+\Mir^2)^2 (q^2+\Mir^2) (k^2+\Mir^2)
          (|\p+\q+\k|^2+\Mir^2)} ,
\label {eq:n2}
\end {eqnarray}
where
\begin {equation}
   \Mir^2 \equiv - 2 m \, \infmu = {4\pi \alpha\over\lambda^2}
\end {equation}
is infinitesimal.
It is not strictly necessary to calculate this term because, by
dimension analysis, the last integral is proportional to
$1/\Mir$ (in three dimensions) and so will only contribute to
the cancellation of IR divergences in (\ref{eq:IRsafe}) and not the
finite remainder.  However, it's reassuring to check the cancellation.
{}From Ref.\ \cite{Braaten&Nieto},%
\footnote{
   See also Ref.\ \cite{rajantie}, which has a useful collection of
   dimensionally regulated three dimensional integrals.
}
\begin {eqnarray}
   \int_{\p\q\k}
   {1\over (p^2+\Mir^2) (q^2+\Mir^2) (k^2+\Mir^2)
          (|\p+\q+\k|^2+\Mir^2)}
\nonumber\\
   = {\Mir\over 16 \pi^3} \left[
        -{1\over 2\eps}
        + {3\over2} \ln \left(2 \Mir\over\bar\Mren\right)
        + \ln 2
        - 2
     \right]
     + O(\eps) .
\end {eqnarray}
Now differentiate with respect to ${\cal M}^2$ to obtain
minus $4$ times the corresponding integral in (\ref{eq:n2}).
Then
\begin {equation}
   \dn_2 = {32\pi a^2\over\lambda^5} \sqrt{\pi\over\alpha} \left[
       - {1\over2\eps}
       + {3\over4}\ln\left(16\pi\alpha\over\bar\Mren^2\lambda^2\right)
       + \ln 2
       - {1\over2}
      \right] .
\label {eq:dn2final}
\end {equation}

Finally, we need the last diagram of Fig.\ \ref{fig:IRsafe} with mass
regularization, corresponding to
\begin {equation}
   \dn_3 = - \Delta\tPi_\sunset(0)\, T \int_\p {1\over (\tomega_p)^2} .
\end {equation}
In dimensional regularization,
\begin {equation}
   \int_\p {1\over (p^2 + \Mir^2)^2}
   = {\Gamma\left(2-{d\over2}\right) \over (4\pi)^{d/2}} \, \Mir^{d-4}
   = {1\over 8\pi\Mir}
             \left[1 - \eps \ln \left(2\Mir\over\bar\Mren\right)\right]
      + O(\eps^2) ,
\label {eq:intsqrm}
\end {equation}
giving
\begin {equation}
   \dn_3 =
   - {\Delta\tPi_\sunset(0)\over\lambda^3 T} \sqrt{\pi\over\alpha}
      \left[1
        - {\eps\over2} \ln \left(16\pi\alpha\over\bar\Mren^2\lambda^2\right)
        + O(\eps^2) \right] .
\label {eq:n3a}
\end {equation}
Because of the $\alpha^{-1/2}$  in this equation, we will need the result
for the mass-regulated $\Delta\tPi_\sunset(0)$ through $O(\sqrt\alpha)$.
The $O(\alpha^0)$ piece, which is independent of the IR regulator, was
calculated in Ref.\ \cite{trap} as part of calculating
$\muc$, and we will rederive it in section \ref{sec:sunset}.  The result is
\begin {equation}
   \Delta\Pi_\sunset(0)
   = - {32\pi a^2 T\over\lambda^2} \left\{
        {1\over 2\eps}
         + \ln(\bar\Mren\lambda)
         + C_1
      \right\} ,
\label {eq:pisunset}
\end {equation}
with $C_1$ as in (\ref{eq:C1}).
To get the $O(\sqrt{\alpha})$ piece in the mass-regulated version, we start
with the integrals corresponding to evaluating Fig. \ref{fig:IRsafemu}
with an infinitesimal chemical potential,
\begin {equation}
   \Delta\tPi_\sunset(0) =
   {1\over2} \left(-{8\pi a\over m}\right)^2 \sumint_{QK}
       {1-\delta_{q_0,0}\delta_{k_0,0}
        \over (i q_0 + \tomega_q)
               (i k_0 + \tomega_k) (i(q_0+k_0)+\tomega_{\q+\k})}
   \,.
\label {eq:sunset1}
\end {equation}
Now differentiate with respect to $\alpha = -\beta\,\infmu$,
\begin {eqnarray}
   {\partial\over\partial\alpha} \, \Delta\tPi_\sunset(0) &=&
   - T \left(-{8\pi a\over m}\right)^2 \biggl[
       \sumint_{QK} {1-\delta_{q_0,0}\delta_{k_0,0}
               \over (i q_0 + \tomega_q)^2
               (i k_0 + \tomega_k) (i(q_0+k_0)+\tomega_{\q+\k})}
\nonumber\\ && \hspace{7em}
       + {1\over2} \sumint_{QK} {1-\delta_{q_0,0}\delta_{k_0,0}
               \over (i q_0 + \tomega_q)
               (i k_0 + \tomega_k) (i(q_0+k_0)+\tomega_{\q+\k})^2}
   \biggr] .
\label {eq:sunset2}
\end {eqnarray}
We want to find the divergent $O(\alpha^{-1/2})$ pieces of this
expression in order to obtain the $O(\sqrt\alpha)$ piece of
(\ref{eq:sunset1}).
The first integral in (\ref{eq:sunset2})
has an IR divergence
associated with $q_0=0$ and $q\to0$.  In this limit, it can be
simplified to
\begin {equation}
   T \int_\q {1\over (\tomega_q)^2}
      \sumint_{K} {1-\delta_{k_0,0}
               \over (i k_0 + \tomega_k)^2} + O(\alpha^0) .
\end {equation}
The infinitesimal chemical potential can be dropped in the $\k$ integral,
since $k_0 \not= 0$ cuts off the infrared, and then the integral is given
by (\ref{eq:intsqr}).  The $\q$ integral is proportional to
(\ref{eq:intsqrm}).  So 
\begin {equation}
   T \int_\q {1\over (\tomega_q)^2}
      \sumint_{K} {1-\delta_{k_0,0}
               \over (i k_0 + \tomega_k)^2}
   = {m^2\,\zeta(\half)\over4\pi\lambda^2\sqrt{\pi\alpha}}
   + O(\sqrt\alpha) + O(\eps) .
\end {equation}
The second integral in (\ref{eq:sunset1}) can be evaluated similarly by
first making the change of variables $Q \to Q-K$, to get
\begin {eqnarray}
&&
   \sumint_{QK} {1-\delta_{q_0,0}\delta_{k_0,0}
               \over (i q_0 + \tomega_q)^2
               (i k_0 + \tomega_k) (i(q_0-k_0)+\tomega_{\q-\k})}
\nonumber\\ && \hspace{10em}
   = T \int_\q {1\over (\tomega_q)^2}
      \sumint_{K} {1-\delta_{k_0,0}
               \over (i k_0 + \tomega_k)(-i k_0 + \tomega_k)} + O(\alpha^0)
\nonumber\\ && \hspace{10em}
   = {m^2\,\zeta(\half)\over 2 \pi\lambda^2\sqrt{\pi\alpha}}
     + O(\alpha^0) + O(\eps).
\end {eqnarray}
Putting it all together to get $\partial\Delta\tPi_\sunset(0)/\partial\alpha$
and then integrating gives
\begin {equation}
   \Delta\tPi_\sunset(0)
   = \Delta\Pi_\sunset(0)
       - {64\pi a^2 T\over\lambda^2} \,\zeta(\half) \sqrt{\alpha\over\pi}
     + O(\alpha) + O(\eps).
\end {equation}
Combining with (\ref{eq:n3a}) and (\ref{eq:pisunset}),
\begin {equation}
   \dn_3
   = {32\pi a^2\over\lambda^5} \biggl\{
        \sqrt{\pi\over\alpha} \left[
          {1\over 2\eps}
          + \ln(\bar\Mren\lambda)
          + C_1
          - {1\over4} \ln\left(16\pi\alpha\over\bar\Mren^2\lambda^2\right)
        \right]
        + 2\,\zeta(\half)
      \biggr\}
   .
\label {eq:dn3final}
\end {equation}

Combining (\ref{eq:dn1final}), (\ref{eq:dn2final}), and (\ref{eq:dn3final})
for the three diagrams of Fig.\ \ref{fig:IRsafe} as
\begin {equation}
   \dnbball = \dn_1 + \dn_2 + \dn_3 ,
\end {equation}
together with
(\ref{eq:C1}) for $C_1$, then yields our final result
(\ref{eq:bballfinal}) for the dimensionally regulated basketball
$\dnbball$.
All the IR divergences cancel, as they should.


\section {Rederivation of \boldmath$\muc$ and \boldmath$\Delta\Pi_\sunset(0)$}
\label {sec:sunset}

Second-order matching results for $\muc$ and $\Delta\Pi_\sunset(0)$ are
derived in Ref.\ \cite{trap}.  However, for the purposes of this paper,
we want them expressed in terms of the same sorts of polylogarithm
integrals that we used in our evaluation of the basketball diagram.
Here, we shall show how to obtain that form.

Start with Fig.\ \ref{fig:IRsafemu} and Eq.\ (\ref{eq:IRsafemu})
for $\Delta\Pi(0)$.  This expression is infrared convergent and so
independent of the choice of IR regulator.  [We will not concern ourselves
here with vanishing corrections, such as the $O(\sqrt\alpha)$ piece that
was important in section \ref{sec:remaining}.]
As we have done before, it will be convenient to nonetheless
introduce an IR regulator
and evaluate the 3+1 dimensional and 3 dimensional pieces separately.
For these particular diagrams, the convenient choice of IR regulator
will be to introduce an infinitesimal chemical potential on just
{\it two}\/ of the three internal propagators:
\begin {eqnarray}
   \Delta\Pi_\sunset(0) &=& -\Pi_\sunset^{(3+1)}(0) + \Pi_\sunset^{(3)}(0)
\nonumber\\
    &=&
    {1\over2} \left(-{8\pi a\over m}\right)^2
      \sumint_{QK} {1-\delta_{q_0,0}\delta_{k_0,0}\over (i q_0 + \omega_q)
               (i k_0 + \tomega_k) (i(q_0+k_0)+\tomega_{\q+\k})} \,.
\end {eqnarray}
The frequency sums in the 3+1 dimensional piece can be evaluated using
standard contour tricks as in Ref.\ \cite{trap} to yield
\begin {eqnarray}
   - \Pi_\sunset^{3+1}(0) &=& 
   {1\over2} \left(8\pi a\over m\right)^2 \int_{\q\k\l}
      n(\omega_q)\,n(\tomega_k)\,n(\tomega_l) \,
      {e^{\beta\tomega_l} - e^{\beta(\omega_q+\tomega_k)}
        \over \tomega_l - \omega_q - \tomega_k}
      \, (2\pi)^d\delta^{(d)}(\l-\q-\k) .
\nonumber \\
   &=&
   {1\over2} \left(8\pi a\over m\right)^2 \int_{\q\k\l}
      \left[ n(\omega_q)\,n(\tomega_k)
             - n(\omega_q)\,n(\tomega_l)
             - n(\tomega_k)\,n(\tomega_l)
             - n(\tomega_l)
      \right]
\nonumber\\ && \hspace{10em} \times
      {(2\pi)^d\delta^{(d)}(\l-\q-\k) \over \omega_l-\omega_q-\omega_k} \,.
\label {eq:Pi1}
\end {eqnarray}
Note that the infinitesimal chemical potential $\infmu$
cancels out in the denominator $\tomega_l-\omega_q-\tomega_k$.
This fact will simplify our analysis later on, and it is the reason that
we chose to put regulator masses on only two of the internal lines rather
than all three.
Now apply a redundant principal part ($\PP$) prescription
to this denominator so that we can separately evaluate the
integrals of each term.  The first term in (\ref{eq:Pi1}) vanishes
on angular integration.
The last term in (\ref{eq:Pi1}), involving just one factor of $n$,
is $O(\eps)$ in dimensional regularization for
the same reasons discussed in Ref.\ \cite{trap}, which are that
\begin {equation}
   \int_{\q\k} \PP
       {(2\pi)^d\delta^{(d)}(\l-\q-\k)\over\omega_l-\omega_q-\omega_k}
   = {\cal O}(\eps)
\end {equation}
and that the remaining $\l$ integration is convergent and cannot generate a
compensating $1/\eps$.
Exchanging integration variables in the second term,
we are then left with
\begin {equation}
   - \Pi_\sunset^{(3+1)}(0) = 
   - {1\over2} \left(8\pi a\over m\right)^2 \int_{\q\k\l}
      \left[ n(\omega^+_q) + n(\omega_q)\right] n(\omega^+_l) \,
      \PP
      {(2\pi)^3\delta^{(3)}(\l-\q-\k) \over \omega_l-\omega_q-\omega_k}
   + O(\eps) .
\end {equation}
This integral has no UV divergences, and so we can set $d=3$ in the
integral.
Now expand the Bose distribution functions as a series in the regulator
fugacity $z$,
\begin {equation}
   n(\omega_p-\mu) = \sum_{a=1}^\infty z^a e^{-a\beta\omega_p} ,
\label {eq:boseexpand}
\end {equation}
giving
\begin {equation}
   - \Pi_\sunset^{(3+1)}(0) = 
   - {1\over2} \left(8\pi a\over m\right)^2
      \sum_{ab} (z^{a+b}+z^b) \int_{\q\k\l}
      e^{-a \beta \omega_q} e^{-b \beta \omega_l}
      \PP
      {(2\pi)^3\delta^{(3)}(\l-\q-\k) \over \omega_l-\omega_q-\omega_k}
   + O(\eps) .
\end {equation}
Rescaling integration variables to be dimensionless,
\begin {equation}
   - \Pi_\sunset^{(3+1)}(0) = 
   - {8 a^2 T\over\lambda^2} \sum_{ab} (z^{a+b}+z^b) I_{ab} + O(\eps) ,
\end {equation}
where
\begin {equation}
   I_{ab} \equiv
   (2\pi)^3 \int_{\q\k\l}
      e^{-a q^2/2} e^{-b l^2/2}
      \PP
      {(2\pi)^3\delta^{(3)}(\l-\q-\k) \over \half(l^2-q^2-k^2)}
   \, .
\end {equation}
Using the methods of Appendix A of Ref.\ \cite{trap}, one may evaluate this
integral, obtaining
\begin {equation}
   I_{ab} = - {1\over (a+b)\sqrt{ab}} \, .
\end {equation}

We can now extract the IR divergences of the sum using the same method as in
section \ref{sec:dn1}:
\begin {equation}
   \sum_{ab} z^{a+b} I_{ab}
   = - \int_0^1 {dt\over t} \left[\Li_{1/2}(zt)\right]^2
   = \pi \ln\alpha
     - 4 \pi K_1
     + O(\sqrt\alpha) ,
\end {equation}
\begin {equation}
   \sum_{ab} z^b I_{ab}
   = - \int_0^1 {dt\over t} \, \Li_{1/2}(t) \, \Li_{1/2}(zt)
   = \pi \ln\left(\alpha\over 4\right)
     - 4\pi K_1
     + O(\sqrt\alpha) ,
\end {equation}
So
\begin {equation}
   - \Pi_\sunset^{(3+1)}(0) = 
   {16 \pi a^2 T\over\lambda^2} \left\{
       \ln\!\left(\alpha\over 2\right)
     - 4 K_1
     \right\}
     + O(\sqrt\alpha) + O(\eps) .
\end {equation}

The corresponding result in the 3-dimensional theory is
\begin {equation}
   - \Pi_\sunset^{(3)}(0) =
   - {512 \pi^3 a^2 T\over\lambda^2} J(\Mir) ,
\end {equation}
where \cite{3dsunset}
\begin {equation}
   J(\Mir) \equiv \int_{\q\k} {1\over q^2(k^2+\Mir^2)(|\q+\k|^2+\Mir^2)}
   = {1\over (4\pi)^2} \left[
        {1\over2\eps}
        + \ln {\bar\Mren\over2\Mir}
        + {1\over2}
     \right]
     + O(\eps) .
\end {equation}
Putting everything together,
one obtains the previously quoted result (\ref{eq:pisunset})
for $\Delta\Pi_\sunset(0)$, with the constant $C_1$
given by (\ref{eq:C1}).
This new form of $C_1$, which is equal to the value derived in
Ref.\ \cite{trap}, can then also be used in the result of Ref.\ \cite{trap}
for $\muc$, which we quoted in (\ref{eq:muc}).


\section {The second-order logarithm for arbitrary \boldmath$N$}
\label{sec:largeN}

As mentioned earlier, Holzmann, Baym, and Lalo\"e argued for the existence
of the logarithmic term at second order.  In order to make their general
argument more concrete, they also presented an approximate large $N$
calculation of the coefficient $c_2'$ of that logarithm. 
It's interesting to compare exact results for the coefficient $c_2'$
to their approximate large $N$ calculation.  For this reason, let us
consider generalizing our 3+1 dimensional theory (\ref{eq:SI}) to a
theory with $\Nc$ complex fields with U($\Nc$) symmetry:
\begin {equation}
   \SI = \int_0^{\beta} d\tau \int d^3x \left[
        \sum_i \psi_i^* \left(
        \partial_\tau - {1\over 2m} \, \nabla^2
        - \mu \right) \psi_i
    + {2\pi a\over m} \, \left(\sum_i \psi_i^* \psi_i\right)^2
    \right] .
\end {equation}
The corresponding 3 dimensional theory can be considered a theory of
$N{=}2\Nc$ real fields and has $O(N)$ symmetry.  The action is again
(\ref{eq:S3}), with $u$ as before, but now
$\phi^2=\phi_1^2+\cdots+\phi_N^2$.
For $\Nc \sim 1$, we reviewed before that the dimensionless cost of each
order of perturbation theory gets a contribution of order
$u/p$ from physics at momentum scale $p$.  For large $\Nc$, the
contribution is order $\Nc u/p$.  (See Refs. \cite{logs,largeN} for
a discussion of large $N$.)  In both cases, the
momentum scale of non-perturbative physics can therefore be characterized
as order $\Nc u$, and the
condition for the theory to be perturbative at the scale $p \sim \lambda^{-1}$
associated with non-zero Matsubara modes is $\Nc u \lambda \ll 1$.
This generalizes the previous condition (\ref{eq:heirarchy}) for
useful expansions of $\Tc(n)$ and the applicability of perturbative
matching to
\begin {equation}
   \Nc a \ll \lambda
\end {equation}
at the transition.  We shall assume this in what follows.

The second-order logarithm in our calculation of $\nc(T)$ arose from
the second diagram of Fig.\ \ref{fig:f3main}, which is proportional to
$\muc$ at the transition.  More specifically, it comes from
the sunset diagram $\Delta\Pi_\sunset(0)$ contribution to $\muc$
(\ref{eq:muc}).  (The sunset diagram is depicted in Fig.\ \ref{fig:IRsafemu}.)
The value of the sunset diagram for general $\Nc$ is simply the value
for $\Nc=1$ multiplied by
\begin {equation}
   {\Nc+1\over2} = {N+2\over4} .
\end {equation}
So, ignoring non-logarithmic second order corrections,
Eq.\ (\ref{eq:muc}) for $\muc$
is modified to
\begin {equation}
   \beta\muc =
   \beta\mucone
   + {(\Nc+1)\over 2} \, {32\pi a^2\over\lambda^2}
      \ln(\bar\Mren \lambda) 
   + \cdots .
\label {eq:mucN}
\end {equation}
Now recall that for the case $\Nc=1$ analyzed in the rest of this paper,
we chose the renormalization momentum scale $\bar\Mren$ to be of order
the momentum scale $u$ for non-perturbative physics.
The reason goes back to section \ref{sec:matchA}: such a choice makes
the critical value of $r_\c(\bar\Mren)$ proportional to $u^2$ by dimensional
analysis.  That meant that the the $r_\c(\bar M)/u^2$ term in the formula
(\ref{eq:muc}) for $\muc$ really gives a straight
${\cal O}(a^2)$ correction without any logarithmic enhancements.
So, if we want to remove the possibility of implicit logarithms in the
$r_\c$ term hiding in the $\cdots$ in (\ref{eq:mucN}), we should again
choose $\bar\Mren$ of order the momentum scale of non-perturbative physics,
which in the present context is
\begin {equation}
   \bar\Mren \sim \Nc u .
\end {equation}
So
\begin {equation}
   \beta\muc =
   \beta\mucone
   + {(\Nc+1)\over 2} \, {32\pi a^2\over\lambda^2} \,
      \ln\!\left(\Nc a\over\lambda\right) 
   + O\left(\Nc a^2 \over \lambda^2\right) ,
\end {equation}
where, for this section only, the notation
$O(\Nc a^2/\lambda^2)$ is meant to assert that
there are no additional factors of logarithms.

The second diagram of Fig.\ \ref{fig:f3main} gives a factor of $\Nc$
as well as the factor of $\muc$.  The ideal gas result $n_0(T)$,
depicted by the first diagram, also has a factor of $\Nc$.
The generalization of (\ref{eq:ncexpand}) for $\nc(T)$ is
\begin {equation}
   \nc(T) = \Nc \lambda^{-3} \left\{
      b_0
      + b_1 a\lambda^{-1}
      + {(\Nc+1)\over2}\,b_2' a^2\lambda^{-2} [\ln(\Nc a\lambda^{-1}) + b_2'']
      + O(\Nc a^2\lambda^{-2})
      \right\} ,
\end {equation}
where $b_0$, $b_1$, and $b_2'$ are as in (\ref{eq:bi}) but with
\begin {equation}
   \kappa \equiv {\Delta\phiphic \over \Nc u} .
\end {equation}
The solution for $\Tc(n)$ can be written in the form
\begin {equation}
  {\Delta\Tc\over T_0} =
     - {128\pi^3\kappa\over \zeta(\threehalf)} \, {a\over\lambda_0}
     + A \, {a^2\over\lambda_0^2} \ln\left(\Nc a\over\lambda_0\right)
     + O\left(a^2\over\lambda_0^2\right) ,
\end {equation}
where
\begin {equation}
   A = - (\Nc+1) \, {32\pi\,\zeta(\half)\over 3\,\zeta(\threehalf)}
\label {eq:Aexact}
\end {equation}
and
$\lambda_0 \equiv \lambda(T_0)$.
In the large $N$ limit,
$\kappa \to -1/96\pi^2$ \cite{largeN} and
the coefficient $A$ of
the logarithm becomes
\begin {equation}
   A \to - \Nc \, {32\pi\,\zeta(\half)\over 3\,\zeta(\threehalf)}
      \simeq 18.7327 \, \Nc .
\label {eq:exactN}
\end {equation}
This is the exact large $N$ result for this coefficient
if large $N$ is defined to mean
taking a large number of
fields in the original 3+1 dimensional theory, as we have
above.  For comparison, the calculation of Ref.\ \cite{logs} was
an approximate large $N$
calculation made in the O($N$) three-dimensional theory
using a rough physically-motivated
UV momentum cut-off of $\Lambda \approx (2\pi)^{1/2}/\lambda$.  Their
approximate result for $A$, expressed in terms of $\Nc=2N$, was
\begin {equation}
   A_{\mbox{\scriptsize (Ref.\ \cite{logs})}}
     \to \Nc \, {256\pi\over 3\,\zeta(\threehalf)  \, \lambda\Lambda}
     \approx \Nc \, {256\pi\over 3\,\zeta(\threehalf) \, (2\pi)^{1/2}}
     \simeq 40.9 \, \Nc .
\label {eq:Aholz}
\end {equation}
This differs by roughly a factor of two from (\ref{eq:exactN}).
Amusingly, the approximation (\ref{eq:Aholz})
of Ref.\ \cite{logs} does well if
naively applied to $\Nc=1$, giving
$A_{\mbox{\scriptsize (Ref.\ \cite{logs})}} \approx 40.9$.
In contrast, the exact answer (\ref{eq:Aexact}) derived in this paper is
$A \simeq 37.4565$.


\section* {ACKNOWLEDGMENTS}

This work was supported by the U.S. Department
of Energy under Grant Nos.\ DE-FG03-96ER40956
and DE-FG02-97ER41027.

\appendix


\section {Some dimensionally regulated integrals}
\label {app:ints}

In this appendix, we show how to do basic single-momentum integrals used
in the main text, such as
(\ref{eq:int2}), (\ref{eq:intsqr}), and (\ref{eq:basicint}).
We will do the $p_0$ frequency sums first, using standard contour tricks.
One can just as easily get the same results by doing the $\p$
integrations first (though one must be careful in that case about cuts).

Start with the basic integral
\begin {equation}
   \sumint_P {1\over ip_0+\omega_p-\mu}
      = \int_\p [n(\omega_p-\mu) + \half].
\end {equation}
The $+\half$ comes from the contour integration
at infinity but can be ignored, since
its integral vanishes in dimensional regularization.
The Bose distribution function can then
be expanded as in (\ref{eq:boseexpand}).
If each term is integrated in $d{=}3{-}\eps$ dimensions,
using the $t=0$ case of
\begin {equation}
   \int {d^d p\over (2\pi)^d} \> p^t e^{- \lambda p^2}
   = {\lambda^{-(d+t)/2} \Gamma\left(d+t\over2\right)
      \over(4\pi)^{d/2} \Gamma\left(d\over2\right)} \,,
\label {eq:moosepoop}
\end {equation}
one obtains
\begin {equation}
   \sumint_P {1\over ip_0+\omega_p-\mu}
   = \Mren^\eps \lambda^{-d} \, \Li_{d\over2}(z) ,
\label {eq:basicz}
\end {equation}
where $z=\exp(\beta\mu)$.
If we set $\mu=0$ in dimensions where the result will still converge
(and then analytically continue to other dimensions), we get the
integral (\ref{eq:basicint}) used in the main text.
If we first differentiate with respect to $\mu$ and then set $\mu=0$,
we get (\ref{eq:intsqrfull}) [which, as discussed in the text, is
equivalent to (\ref{eq:intsqr}) in dimensional regularization].

We can do the integral (\ref{eq:int2}) of the main text as
\begin {equation}
   \sumint_P {1\over(ip_0+\omega_p)(-ip_0+\omega_p)} =
   \sumint_P {1\over2\omega_p}
      \left[{1\over i p_0 + \omega_p} + {1\over -i p_0 +\omega_p}\right]
   = \int_\p {n(\omega_p)+\half\over\omega_p}
   = \int_\p {n(\omega_p)\over\omega_p}
   \, .
\end {equation}
The last integral can also be done by expanding $n$ as above, using the
$t{=}-2$ case of (\ref{eq:moosepoop}), with the result
quoted in the text.


\end {document}